\documentclass[useAMS,usenatbib]{mn2e}
\usepackage{graphics,graphicx,epsfig,psfig}
\usepackage[normalem]{ulem}
\usepackage[dvipsnames]{color}
\usepackage{soul,xcolor}
\usepackage{amsmath, amssymb}
\usepackage[]{inputenc,amssymb}
\usepackage{booktabs,caption}
\usepackage{amsmath}
\usepackage{multirow}

\def \tcg{\textcolor{webgreen}}

\setstcolor{red}

\def \be{\begin{equation}}
\def \ee{\end{equation}}
\def \bea{\begin{eqnarray}}
\def \eea{\end{eqnarray}}
\def \etal{{et al.}}

\definecolor{webgreen}{rgb}{0,.5,0}
\definecolor{webbrown}{rgb}{.6,0,0}

\usepackage[%
    colorlinks = true,%
    linkcolor = blue,%
    urlcolor  = blue,%
    citecolor = webgreen,%
    anchorcolor = blue]{hyperref}

\newcommand{\ufhref}[3][blue]{\href{#2}{\color{#1}{#3}}}%

\title[Effect of cosmic rays on superbubbles]{Lack of thermal energy in superbubbles: hint of cosmic rays?}
\author[Gupta et al.]
{Siddhartha Gupta$^{1,2}$ \thanks{siddhartha@rri.res.in}, Biman B. Nath$^1$, Prateek Sharma$^2$, David Eichler$^3$\\
$^1$Raman Research Institute, Sadashiva Nagar, Bangalore 560080, India\\
$^2$Joint Astronomy Programme and Department of Physics, Indian Institute of Science, Bangalore 560012, India\\
$^3$Dept. of Physics, Ben-Gurion University, Be'er-Sheba 84105, Israel}
\begin{document}
\maketitle
\label{firstpage}
\begin{abstract}
Using analytic methods and $1$-D two-fluid simulations, we study the effect of cosmic rays (CRs) on the dynamics of interstellar superbubbles (ISBs) driven by multiple supernovae (SNe)/stellar winds in OB associations. In addition to CR advection and diffusion, our models include thermal conduction and radiative cooling. We find that CR injection at the reverse shock or within a central wind-driving region can affect the  thermal profiles of ISBs and hence their X-ray properties. Even if a small fraction ($10-20\%$) of the total mechanical power is injected into CRs, a significant fraction of the ram pressure at the reverse shock can be transferred to CRs. The energy transfer becomes efficient if (1) the reverse shock gas Mach number exceeds a critical value ($M_{\rm th}\gtrsim 12$) and (2) the CR acceleration time scale $\tau_{\rm acc}\sim \kappa_{\rm cr}/v^2$ is  shorter than the dynamical time, where $\kappa_{\rm cr}$ is CR diffusion constant and $v$ is the upstream velocity. We show that CR affected bubbles can exhibit a volume averaged hot gas temperature $1-5\times10^{6}$ K, lower by a factor of $2-10$ than without CRs. Thus CRs can potentially solve the long-standing problem of the observed low ISB temperatures.
\end{abstract}
\section{Introduction}
\label{sec:intro}
Superbubbles driven by stellar winds and supernovae from OB associations are the instruments of stellar feedback that regulate galaxy evolution. These expanding shells form the crucial link between stars and the interstellar medium (ISM) by depositing thermal and kinetic energy, and thereby influencing the star formation process. On a larger scale, they can launch galactic scale outflows if certain conditions are fulfilled (\citealt{Nath2013}; \citealt{Sharma2014}). 

The classical model of \citet{Weaver1977} provided the basic framework for wind driven bubbles. They described the shock structure expected in ISBs, and worked out the dynamics in the self-similar phase of evolution. 
\citet{MacLow1988} proposed an ISB model by considering  correlated SNe and discussed its effects on the galactic scale. All these ideas have been studied in detail with numerical simulations (\citealt{Keller2014}; \citealt{Yadav2017}; \citealt{Vasiliev2017}).

In recent years, X-ray observations of various bubbles have led to a closer look into their dynamics (\citealt{Townsley2006} and references therein). The presence of a dominant soft X-ray component at $\sim 2\times 10^{6}$ K has also been highlighted in some of these studies (e.g., \citealt{Chu2003a}; \citealt{Maddox2009}). Most of these studies provide the temperature and density of the X-ray emitting plasma which are often used to understand the effective driving force acting on the dense shell (\citealt{Pellegrini2011}; \citealt{Lopez2011}; \citealt{Lopez2014}). However, these analyses show that the best-fit X-ray luminosity and temperature are often lower than that expected from the classical bubble model (\citealt{Chu2003b}; \citealt{Harper2009}). Even with thermal conduction, which makes the bubble denser and cooler, the discrepancy is not fully resolved. Therefore, we investigate the role of CRs in modifying the ISB properties.

Supernova remnants also show results similar to ISBs. \citet{Chevalier1983} re-examined the blast wave solution to model the effect of relativistic particles (see also \citealt{Vink2010}; \citealt{Bell2014}). He showed that the injection of CRs can reduce the thermal energy inside the blast wave. This idea has been confirmed by analyzing the post shock temperature in RCW $86$ supernova remnant (\citealt{Helder2009}).

Recently, It has been suggested that ISBs are the preferred sites of CR acceleration instead of isolated supernova remnants (SNRs). Since massive stars usually form in clusters, most of the power injected by supernovae into the ISM is mediated through superbubbles and not isolated SNRs (\citealt{Higdon2005}). \citet{Binns2005} have suggested that the isotopic anomalies in the composition of Galactic cosmic rays, in particular, the enhanced $^{22}Ne/^{20}Ne$ ratio, are suggestive of CRs being accelerated out of the matter inside ISBs. \citet{Parizot1999} argued that a superbubble model for the origin of Galactic cosmic rays can explain the evolution of light elements Li, Be and B. Gamma-ray observations of the Cygnus superbubble have also shown that CRs are accelerated in ISBs (\citealt{Ackermann2011}). Recently \citet{Eichler2017} has suggested that cosmic ray grammage traversed is correlated with the properties of the source, meaning that the escape occurs near the production site, suggestive of ISBs. Observations of extragalactic superbubbles in IC 10 ({\citealt{Heesen2014}) and the Large Magellanic Cloud (\citealt{Butt2008}) have also  suggested the ISB origin of CRs. Therefore the dynamics of ISBs, and the possible effects of CR on them, deserve to be studied in detail. 
 
The role of CR feedback in galaxy evolution has also been discussed by several authors (e.g., \citealt{Salem2013}; \citealt{Booth2013}; \citealt{Wiener2017}). Although these studies included various physical processes e.g. self-gravity, cooling, star formation, the effects of the individual processes are difficult to disentangle.

In this paper, we present a two-fluid model of ISB. We start with a standard ISB model and include thermal conduction, radiative cooling and CR diffusion one by one to understand the role of each processes. We show that in the absence of CRs the thermal pressure of hot gas  depends on the ambient density (almost independent of density profile) and the shell speed. Then we explore the effect of CRs. We have found that CRs can affect ISB via shock interactions and their effects mainly depend on the shock Mach number (\citealt{Drury1981}; \citealt{Becker2001}) and CR diffusion coefficient.

The contents of this paper are organized as follows. We present a broad analytical frame work in section \ref{sec:analytical}. The details of the simulation set-up are given in section \ref{sec:setup}. The results from various runs are presented in section \ref{sec:result}.  In section \ref{sec:discussion} we discuss the dependence of the results on  various parameters. In section \ref{sec:anaobs} we comment on the astrophysical implication of this study. Finally, we conclude in section \ref{sec:summary} by highlighting the main results of this paper.
\section{Analytical prelude}
\label{sec:analytical}
Consider the powerful wind from OB association that drives an ISB. The dynamics and the structure of an ISB is usually understood by the momentum and energy conservation equations:
\begin{eqnarray}
\label{eq:m1}
\frac{d}{dt}(M\,\dot{R}) = 4\pi R^{2}\,P
\end{eqnarray}
\begin{eqnarray}
\label{eq:e1}
\frac{d}{dt}\left(\frac{4}{3}\pi R^3 \frac{P}{\gamma-1}\right) = L_{\rm w}- 4\pi R^{2}\,\dot{R}\, P - L_{\rm loss} \ ,
\end{eqnarray}
where $P$ is the pressure inside the bubble (assumed to be $\gg$ ambient pressure), $R$ is the position of the swept-up ISM (hereafter `shell') w.r.t. the central source\footnote{By the source region, we mean a region within which most of the stars are located, i.e., the region which is driving the ISB.}, $M=\int_{0}^R dr\, 4 \,\pi\,r^2\,\rho(r)$ is the swept-up ambient  mass, $\rho(r)$ is the ambient density (c.f. equation (\ref{eq:rhof})), $L_{\rm w}$ is the wind power, and  $L_{\rm loss}$ is the loss of energy due to radiative cooling.

In order to find a general solution of equations (\ref{eq:m1}) - (\ref{eq:e1}), we choose the ambient density profile as
\begin{eqnarray}
\label{eq:rhof}
\rho(r) = \rho_{\rm c} \left(\frac{r_{c}}{r}\right)^s\, ,
\end{eqnarray}
where the choice of the parameter `$s$' determines the ambient density profile. In next sections, we discuss the solutions in different cases.
\subsection{One-fluid standard ISB}
\label{subsec:onefluidtheory}
\subsubsection{Adiabatic evolution}
\begin{table}
\caption{The constant factors presented in equations (\ref{eq:c1c2}) and (\ref{eq:prs_anafit}).}
\begin{center}
\begin{tabular}{c c c c c c c}
  \hline\hline
  $\gamma$ & $s$ & $C_{\rm 1}$ & $C_{\rm 2}$ & $\xi$\\
    \hline
    \multirow{3}{*}{5/3}
 & $0.0$ & $0.763$ & $0.163$ & $0.78$ \\ 
 & $1.0$ & $0.607$ & $0.285$ & $0.83$ \\ 
 & $2.0$ & $0.376$ & $1.000$ & $1.00$ \\
  \hline
    \multirow{3}{*}{13/9}
 & $0.0$ & $0.732$ & $0.150$ & $0.78$ \\ 
 & $1.0$ & $0.580$ & $0.272$ & $0.83$ \\ 
 & $2.0$ & $0.357$ & $1.000$ & $1.00$ \\
 \hline
    \multirow{3}{*}{4/3}
 & $0.0$ & $0.708$ & $0.140$ & $0.78$ \\ 
 & $1.0$ & $0.558$ & $0.262$ & $0.83$ \\ 
 & $2.0$ & $0.341$ & $1.000$ & $1.00$ \\
    \hline
\end{tabular}
\end{center}
\label{tab:const}
\end{table}
Consider the case when the dynamical time ($t_{\rm dyn}$) of an ISB is shorter than the cooling time-scales of its different regions (for  details see \citealt{Castor1975}, sections $3$ and $5.2$ in \citealt{Gupta2016}). Under this assumption, the contribution of $L_{\rm loss}$ in the r.h.s of equation (\ref{eq:e1}) is negligible. Assuming that at any given instant, $R\propto t^{\alpha}_{\rm dyn}$ (where $\alpha > 0$) and substituting $P$ from equation (\ref{eq:m1}) to equation (\ref{eq:e1}), we obtain
\begin{eqnarray}
\label{eq:Rgen}
R =  C_{\rm 1}\ L_{\rm w}^{1/(5-s)}\,(\rho_{\rm c}r^s_{\rm c})^{-1/(5-s)}\,t_{\rm dyn}^{3/(5-s)}
\end{eqnarray}
\begin{eqnarray}
\label{eq:Pgen}
P = C_{\rm 2}\ L_{\rm w}^{(2-s)/(5-s)}\,(\rho_{\rm c}r^s_{\rm c})^{3/(5-s)}\,t_{\rm dyn}^{-(4+s)/(5-s)}
\end{eqnarray}
where 
\begin{eqnarray}
\label{eq:c1c2}
C_{\rm 1} & = & \left[\frac{(\gamma-1)\,(5-s)^3\,(3-s)}{4 \pi \{(63-18 s)\gamma+s(2s+1)-28\}}\right]^{1/(5-s)} \ {\rm and} \nonumber \\
C_{\rm 2} & = & \left[\frac{21-6s}{(5-s)^2(3-s)}\right]C_{\rm 1}^{(2-s)}
\end{eqnarray}
Note that, the choice $s=0$ corresponds to an ISB expanding in a uniform medium (\citealt{Weaver1977}). 

The pressure $P$ (the shocked wind (SW) pressure) given in equation (\ref{eq:Pgen}) can be re-written in following form, 
\begin{eqnarray}
\label{eq:prs_anafit}
P=\xi\,  \rho \, v_{\rm sh}^2 \, , 
\end{eqnarray}
where $v_{\rm sh}$ is the shell velocity ($=dR/dt$; see equation (\ref{eq:Rgen})), $\rho$ is the unshocked ISM density (i.e., the upstream density; equation (\ref{eq:rhof})) and $\xi=C_{\rm 1}^{(s-2)}\,C_{\rm 2} \left[(5-s)/3\right]^2$. The values of $C_{\rm 1}$, $C_{\rm 2}$ and $\xi$ for different adiabatic constants ($\gamma$)  and $s$ are given in Table \ref{tab:const}. Note that, $\xi$ depends weakly on the density profile ($s$), and therefore,  equation (\ref{eq:prs_anafit}) is a robust estimate of the interior pressure of an ISB.

Using equation (\ref{eq:prs_anafit}), the position of the reverse shock can be readily obtained by equating the ram pressure and the gas pressure of the SW region. This gives,
\begin{eqnarray}
\label{eq:Rrs}
R_{\rm rs} =  \left(\frac{1}{4\pi \xi}\right)^{1/2}\,\rho^{-1/2}\,\dot{M}^{1/2}\,v_{\rm w}^{1/2}\, v_{\rm sh}^{-1}\ ,
\end{eqnarray}
where $\dot{M}$ is the mass loss rate of the source, and $v_{\rm w}\approx (2L_{\rm w}/\dot{M})^{1/2}$ (\tcg{Chevalier \& Clegg 1985}, hereafter \citealt{CC1985}) is the wind velocity. For a uniform ambient medium, the position of the reverse shock is given by,
\begin{eqnarray}
\label{eq:Rrs_uni}
R_{\rm rs} = 2.45\ \rho_{\rm 2}^{-3/10} \,\dot{M}_{\rm -4}^{1/2}\, L_{\rm39}^{-1/5}\, v^{1/2}_{\rm3}\,\,t^{2/5}_{6} \ {\rm pc} \,.
\end{eqnarray}
Here $\rho_{\rm 2}=\rho/(10^2\,m{\rm_H\,cm^{-3}})$, $\dot{M}_{\rm -4}=\dot{M}/(10^{-4}\,\rm{M_{\odot}\,yr^{-1}})$,  $L_{\rm 39}=L_{\rm w}/(10^{39}\,\rm{erg\,s^{-1}})$,  $v_{\rm 3}=v_{\rm w}/(10^{3}\,\rm{km\,s^{-1}})$ and $t_{\rm 6}=t_{\rm dyn}/(1\,\rm{Myr})$.
Neglecting thermal conduction, the density of the shocked wind region ($\rho_{\rm sw}$) can be estimated by applying the shock jump condition at the reverse shock. This gives 
\begin{eqnarray}
\label{eq:rhosw}
\rho_{\rm sw} = \left(\frac{\gamma+1}{\gamma-1}\right)\,\xi\,\rho\, v_{\rm sh}^2\,v_{\rm w}^{-2}
\end{eqnarray}
\begin{eqnarray}
\label{eq:Tsw}
T_{\rm sw} = \frac{2(\gamma-1)}{(\gamma+1)^2}\frac{\mu m_{\rm H}}{k_{\rm B}}v^2_{\rm w}
\end{eqnarray}
If one uses a typical wind velocity $v_{\rm w}\sim 2000\ {\rm km\, s^{-1}}$ (\citealt{Leitherer1999}), then $T_{\rm sw}\approx 5\times 10^{7}$ K, which is higher than the observed ISB temperatures. At these temperatures and densities, the electron and ion temperatures are equalized by Coulomb collisions.
\subsubsection{Thermal conduction}
\label{subsubsec:tc}
Thermal conduction transfers heat from the hot bubble interior to the cooled shell, which causes the evaporation of the swept-up shell mass into the interior of the bubble (\citealt{Cowie1977}). The detailed self-similar analysis of \citet{Weaver1977} showed that for a uniform ambient medium, thermal conduction decreases the temperature of the hot gas and enhances the density of the interior. Choosing the classical isotropic thermal conductivity $\kappa_{\rm th}=6\times 10^{-7} T^{5/2}$ cgs (\citealt{MacLow1988}), we have,
\begin{eqnarray}
\label{eq:rhosw_tc}
\rho_{\rm sw,tc} \simeq 0.27\,L_{\rm 39}^{6/35}\rho_{2}^{19/35} t_{6}^{-22/35}\left[1-R_{\rm rs}/R_{\rm cd}\right]^{2/5} m{\rm_H\,cm^{-3}}
\end{eqnarray}
\begin{eqnarray}
\label{eq:Tsw_tc}
T_{\rm sw,tc} \simeq 1.26\times 10^{7} L_{\rm 39}^{8/35}\rho_{2}^{2/35}\,t_{6}^{-6/35}\,\left[1-R_{\rm rs}/R_{\rm cd}\right]^{-2/5}\, {\rm K}
\end{eqnarray}

The use of classical  thermal conduction, however, can be questioned because the assumption of isotropy is not valid in the presence of magnetic field. Furthermore, at the early times, the mean free path of electrons ($\lambda_{\rm m}$) $\gtrsim$ the temperature gradient scale ($l_{\rm T}$), which can cause conduction to be saturated. Therefore, equations (\ref{eq:rhosw_tc}) and (\ref{eq:Tsw_tc}) represent the upper limiting case of thermal conduction.
\subsubsection{Radiative cooling}
\label{subsubsec:coolana}
For a radiative bubble, cooling can delay the formation of the reverse shock (see section 5.2 in \citealt{Gupta2016}). However, once it forms, the qualitative description remains the same as above but their quantitative results change because of the term $L_{\rm loss}$ (see equation (\ref{eq:e1})). The cooling loss rate ($L_{\rm loss}$) is defined as 
\begin{eqnarray}
L_{\rm loss}=\int_{0}^R dr\, 4 \,\pi\,R^2 n_{\rm e}n_{\rm i}\Lambda_{\rm N}\ ,
\end{eqnarray}
where $n_{\rm i}/n_{\rm e}$ is the electron/ion number density, and $\Lambda_{\rm N}$ is the normalized (w.r.t. ($n_{\rm i}n_{\rm e})$) cooling rate. 

Note that, although the contact discontinuity (hereafter `CD') occupies a much smaller volume compared to the size of the bubble, the contribution of the term $L_{\rm loss}$ mainly comes from the CD, since this is the region where the temperature passes though the peak of the cooling curve. It can be shown that, if $t_{\rm dyn} $ is longer than the shell cooling time scale ($\tau_{\rm shell}$, see equation ($4$) in \citealt{Gupta2016}) then the term $(L_{\rm w}-L_{\rm loss})$ in equation (\ref{eq:e1}) can be approximated as $\eta L_{\rm w}$ where $\eta$ ($\leq 1$) can be thought of as an energy efficiency parameter. This parameter ($\eta$) depends mainly on the ambient density (see section 6.3 in \citealt{Gupta2016}). 
The expressions in equations (\ref{eq:Rgen}) and (\ref{eq:Pgen}) are accordingly changed by replacing $L_{\rm w}$ by $\eta L_{\rm w}$.
\subsection{Two-fluid ISBs}
\label{subsec:twofluidisb}
The discussions so far have been confined to the case of one-fluid ISB. Here we discuss the modifications due to CRs. 

The effects of CRs depend on three main parameters : (1) the regions where CRs are accelerated/injected, (2) the fraction of energy that goes into CRs and (3) the CR diffusion coefficient. Note that all these parameters are uncertain and therefore we cover a range of models and parameters.

For CR affected ISBs, the interior gas can be considered as a mixture of thermal and non-thermal particles. It is useful to take an effective adiabatic index $\gamma_{\rm cr}(=4/3)<\gamma<\gamma_{\rm th}(=5/3)$ to infer the dynamics. The adiabatic index of the gas mixture depends on the fraction of the thermal and non-thermal (CR) particles in the gas, which may vary between different regions in the ISB. Moreover, the description becomes more complicated when CR diffusion is considered. For a preliminary understanding, we divide the discussion based on the CR injection region.
\subsubsection{Injection at the shocks}
Consider a scenario in which the CRs are accelerated at the forward shock (hereafter FS; equivalent to CR injection at FS). In this case, we do not expect to see any change in the interior because CRs do not penetrate CD. However, if CR acceleration happens at the reverse shock (hereafter RS) then CRs can diminish the thermal pressure in the SW region. It results in a decrease of SW temperature which reduces the effect of thermal conduction. 

For injecting CR at the shock, a parameter is commonly used (\citealt{Chevalier1983}; \citealt{Bell2014}), which is defined as
\begin{eqnarray}
\label{eq:pcrin}
w = \frac{p_{\rm cr}}{p_{\rm th}+p_{\rm cr}}
\end{eqnarray}
where $p_{\rm cr}$ and $p_{\rm th}$ are the CR and thermal pressure. The implications for CR injection at shocks (with and without CR diffusion) are discussed in detail in section \ref{subsubsec:crinjshk}.

\subsubsection{Injection at the source region}
\label{subsubsec:insrc_ana}
CR injection in the source region presents a different scenario. In this case, it is assumed that some fraction of total deposited energy by stellar winds and/or SNe goes into CRs (\citealt{Salem2013}). Here, the injection parameter is defined as
\begin{eqnarray}
\label{eq:ecrin}
\epsilon_{\rm cr} = \frac{E_{\rm cr}}{E_{\rm IN}}\, ,
\end{eqnarray}
where $E_{\rm cr}$ is the energy deposited in CRs and $E_{\rm IN}$ is the total deposited energy. 

The main difference between equations  (\ref{eq:pcrin}) and (\ref{eq:ecrin}) is that the former parameter is defined at a shock whereas the latter one is defined at the source. Unlike the previous case, CR injection in the source region causes a free wind CR pressure profile ($p_{\rm cr}$) which makes the bubble structure different from the standard one-fluid ISB. 

To obtain CR profile in the free wind region, one has to solve a two-fluid steady state model along the lines of \citet{CC1985}. Inside the free wind, we expect that $p_{\rm cr}\propto \rho^{4/3}$ (adiabatic expansion of the wind) i.e. $p_{\rm cr} \propto r^{-8/3}$. The normalization constant depends on the fraction of energy injected as CRs and also on the input source parameters (see Table 1 in \citealt{CC1985}). Our numerical simulation shows that the power law profile, which is valid only in the absence of CR diffusion, and can be written as
\begin{eqnarray}
\label{eq:pcrfree}
\frac{p_{\rm cr}}{\dot{M}^{1/2} L_{\rm w}^{1/2} R_{\rm src}^{-2}} \simeq 0.011\, \epsilon_{\rm cr}\,\left(\frac{r}{R_{\rm src}} \right)^{-8/3}
\end{eqnarray}
where $R_{\rm src}$ is the radius of the source region. 

As the free wind reaches the reverse shock, one does not have a one-fluid shock. Two-fluid shocks have been previously studied by several authors (\citealt{Drury1981}; \citealt{Drury1986}; \citealt{Wagner2007}). CR diffusion plays a key role in two-fluid shock. For some upstream parameters,  CRs can diffuse across the shock and the downstream CR pressure can increase. The downstream CRs then diffuse further, which changes the upstream CR pressure. As a result, the CR particles cross the shock multiple times  (which is also known as diffusive shock acceleration, \citealt{Becker2001}) and CRs can modify the shock.

\citet{Drury1981} showed that, for a two-fluid shock, parameters of the downstream fluid can have three possible solutions: (1) Globally smooth, (2) Discontinuous solution and (3) Gas mediated sub-shock. \citet{Becker2001} classified the solution parameter space based on the upstream gas and CR Mach numbers (see their equation ($25$)), which are defined as $M_{\rm th}=v / a_{\rm th}$ and $M_{\rm cr}=v / a_{\rm cr}$ respectively  where $a_{\rm th}=\sqrt{\gamma_{\rm th} p_{\rm th}/\rho}$ and $a_{\rm cr}=\sqrt{\gamma_{\rm cr} p_{\rm cr}/\rho}$  are the adiabatic sound speed for the respective fluids. They showed that if the gas  Mach number exceeds a critical value $12.3$, then the downstream (post shock) CR pressure dominates over thermal pressure and the shock structure is globally smooth (see their Figures $8$  and $16$). We, therefore, estimate the reverse shock and forward shock Mach numbers to predict the shock structure.

In case of the reverse shock (RS), the upstream wind velocity can be assumed to be the same as the wind velocity ($v_{\rm w}$) because the RS velocity $\ll v_{\rm w}$ (see equation (\ref{eq:Rrs})). The upstream sound speed is estimated using the steady state free wind pressure and density profiles (\citealt{CC1985}), which gives $a_{\rm th,rs}=0.56 (R_{\rm rs}/R_{\rm src})^{-2/3} \dot{M}^{-1/2}\,L_{\rm w}^{1/2}$. For the uniform ambient, this yields
\begin{eqnarray}
\label{eq:Rrs_mach}
M_{\rm th,rs} = \frac{v_{\rm w}}{a_{\rm th,rs}} \simeq 8.15\, \eta^{-2/15} R_{\rm src, pc}^{-2/3}\, \rho_{\rm 2}^{-1/5} \dot{M}_{\rm -4}^{1/6}\, L_{\rm39}^{1/30}\,t^{4/15}_{6}
\end{eqnarray}
The most interesting thing about the RS is that the Mach number increases with time. Therefore, the RS is expected to show a globally smooth profile after a timescale $\tau_{\rm cri}$. For a uniform ambient medium, we obtain
\begin{eqnarray}
\label{eq:tau_cri}
\tau_{\rm cri} \simeq 4.65\ \eta^{1/2}\,R_{\rm src, pc}^{5/2}\,\rho_{\rm 2}^{3/4} \,\dot{M}_{\rm -4}^{-5/8}\, L_{\rm39}^{-1/8}\ {\rm Myr}
\end{eqnarray}
Note that, in equation (\ref{eq:Rrs_mach}) we have included the energy efficiency term $\eta$ (discussed in section \ref{subsubsec:coolana}). This shows that a radiative ISB can satisfy the critical Mach number criterion ( $M_{\rm th,rs}>12.3$) at an earlier time than the adiabatic ISB.

The forward shock (FS) Mach number can easily be found by assuming the ambient medium is at rest. This gives
\begin{eqnarray}
\label{eq:Rfs_mach}
M_{\rm th,fs} = \frac{v_{\rm sh}}{a_{\rm th,fs}} \simeq 10\ v_{\rm sh, 2}\, a^{-1}_{\rm th,1} 
\end{eqnarray}
It is worth mentioning that depending on the ambient temperature, $M_{\rm th,fs}$ can exceed $12.3$. Therefore, the FS can also show a globally smooth shock structure. However, note that the Mach number of the forward 
shock decreases with time.

The important point is that, for a globally smooth solution, the upstream kinetic energy is mostly transferred to the downstream CR pressure. This diminishes the thermal pressure and can change the density and temperature profiles of ISBs. 
\begin{table*}
\caption{Details of the runs.}
\label{tab:sim1}
 \centering
 \begin{tabular}{ l c c c c c c c c c} 
  \hline\hline 
 \centering{$[1]$} &  $[2]$ &$[3]$ & $[4]$ & \multicolumn{3}{c}{$[5]$ Micro-physics} & \multicolumn{2}{c}{$[6]$ Simulation box details}\\
 \cmidrule(l){5-7} \cmidrule(l){8-9}                                     
Model & No. of OB stars & $\rho_{\rm c}$   &  Two-fluid  & \scriptsize{CR diffusion}  & \scriptsize{Therm Conduction} & \scriptsize{Cooling} & \scriptsize{Box size} & \scriptsize{Grids}\\
        &  (N) & \scriptsize{($m{\rm_{H}\,cm^{-3}}$)}  & &  \scriptsize{(crd)}  & \scriptsize{(t)} & \scriptsize{(c)} & \scriptsize{[${\rm r_{min}, r_{max}}$]} & \scriptsize{[${\rm n}$]}   \\   
  \hline
${\rm N3\_d2}$                   & $10^{3}$ & $10^{2}$  & N & N  & N & N & [0.1,400.1]   & [8000]\\
${\rm N3\_d2\_t}$               & $10^{3}$ & $10^{2}$  & N & N  & Y & N & [0.1,300.1]   & [6000]\\
${\rm N3\_d2\_c}$              & $10^{3}$ & $10^{2}$  & N & N  & N & Y & [0.1,300.1]   & [6000]\\
${\rm N3\_d2\_c\_t}$         & $10^{3}$ & $10^{2}$  & N & N  & Y & Y & [0.1,300.1]   & [6000]\\
${\rm N3\_d2\_cr}$             & $10^{3}$ & $10^{2}$  & Y & N  & N & N & [0.1,400.1]   & [8000]\\
${\rm N3\_d2\_crd}$           & $10^{3}$ & $10^{2}$  & Y & Y  & N & N & [0.1,1000.1]  & [20000]\\
${\rm N3\_d2\_crd\_t}$       & $10^{3}$ & $10^{2}$  & Y & Y  & Y & N & [0.1,300.1]  & [6000]\\
${\rm N3\_d2\_crd\_c}$      & $10^{3}$ & $10^{2}$  & Y & Y  & N & Y & [0.1,1000.1]  & [20000]\\
${\rm N3\_d2\_crd\_c\_t}$ & $10^{3}$ & $10^{2}$  & Y & Y  & Y & Y & [0.1,500.1]    & [10000]\\
${\rm N3\_d0\_crd\_c}$     & $10^{3}$ & $1$       & Y & Y  & N & Y & [0.1,1400.1]  & [28000]\\
${\rm N3\_d3\_crd\_c}$     & $10^{3}$ & $10^{3}$ & Y & Y  & N & Y & [0.1,1000.1]  & [20000]\\
${\rm N2\_d2\_crd\_c}$     & $10^{2}$  & $10^{2}$ & Y & Y  & N & Y & [0.1,1000.1]  & [20000]\\
${\rm N4\_d2\_crd\_c}$     & $10^{4}$  & $10^{2}$ & Y & Y  & N & Y & [0.1,1000.1]  & [20000]\\
 \hline
\end{tabular}
\raggedright{Note: The symbol ‘N’ and ‘Y’ denote that the processes are respectively switched off and on. The nomenclature of the models can be illustrated with the help of an example: ${\rm N3\_d2\_crd\_c\_t}$ represents a run where the number of $N_{\rm OB}$ stars is $10^3$, the ambient density is  $100\, m{\rm_H\,cm^{-3}}$ and the evolution has been studied using two-fluid equations (${\rm cr}$) with cosmic ray diffusion (${\rm crd}$), radiative cooling (${\rm c}$) and thermal conduction (${\rm t}$).}
\end{table*}

\section{Simulation set-up}
\label{sec:setup}
In order to study the detailed effects of CRs in ISBs, we have developed a 1-D, two-fluid code that we call {\textsc {TFH}} (standing for Two Fluid Hydrodynamics) code.  {\textsc {TFH}}  solves the following set of equations :
\begin{eqnarray}
\label{eq:mass}
\frac{\partial \rho}{\partial t}+\vec{\nabla}.(\rho\,\vec{v})  =  S_{\rm \rho}
\end{eqnarray}
\begin{eqnarray}
\label{eq:momentum}
 \frac{\partial }{\partial t} (\rho\,\vec{v}) +\vec{\nabla}.(\rho\,\vec{v}\otimes\vec{v}) + \vec{\nabla} (p_{\rm th} + p_{\rm cr}) = 0     
\end{eqnarray}
\begin{eqnarray}
\label{eq:energy}
\frac{\partial}{\partial t} \left(e_{\rm th}+ e_{\rm k}\,\right) +\vec{\nabla}.\left[ \left(e_{\rm th}+\,e_{\rm k}\right)\vec{v}\right]  + \vec{\nabla}.\left[\vec{v}\,(p_{\rm th}+p_{\rm cr})\right] \nonumber \\
  =  p_{\rm cr} \vec{\nabla}.\vec{v}  - \vec{\nabla}. \vec{F}_{\rm tc} + q^{-}_{\rm eff} +  S_{\rm e}
\end{eqnarray}
\begin{eqnarray}
\label{eq:energycr}
 \frac{\partial e_{\rm cr}}{\partial t}  +\vec{\nabla}.\left[ e_{\rm cr}\,\vec{v}\right] = - p_{\rm cr} \vec{\nabla}.\vec{v} - \vec{\nabla}.\vec{F}_{\rm crdiff}+ S_{\rm cr}
\end{eqnarray}\\
Here, $\rho$ is the mass density, $\vec{v}$ is the fluid velocity, $p_{\rm th}/p_{\rm cr}$ is the thermal/CR pressure, $e_{\rm k}=\rho\,v^{2}/2$ is the kinetic energy density, $e_{\rm th}=p_{\rm th}/(\gamma_{\rm th}-1)$ and $e_{\rm cr}=p_{\rm  cr}/(\gamma_{\rm cr}-1)$ are the thermal (gas) and CR energy densities respectively (\citealt{Drury1981}; \citealt{Wagner2007}; \citealt{Pfrommer2017}). The terms $S_{\rm \rho}$, $S_{\rm th}$ and $S_{\rm cr}$ in equations (\ref{eq:mass}), (\ref{eq:energy}) and (\ref{eq:energycr}) represent mass and energy terms deposited by the driving source.  $F_{\rm tc}$ and $F_{\rm crdiff}$ represent thermal conduction flux and CR diffusion flux respectively. The term $q^{-}_{\rm eff}$ accounts for the radiative energy loss of the thermal gas. For simplicity, we exclude the gas heating due to CR streaming (\citealt{Guo2008}).

Currently {\textsc {TFH}} has two solvers: (1) a ZEUS-like solver (\citealt{Stone1992}) which performs transport and source terms separately (hereafter, TS; for details, see chapter $5$ in \citealt{Dullemond2009}\footnote{See http://www.mpia.de/homes/dullemon/lectures/fluiddynamics}) and (2) HLL solver (\citealt{Toro1994}; for details, see chapter $10$ in \citealt{Toro2009}). Both solvers use the finite volume method and are first order accurate. Because of the first order scheme, all runs are done with very high resolution (typical resolution is $0.05$ pc). Since {\textsc {TFH}} performs two-fluid simulation, we have defined an effective sound speed as $c_{\rm s, eff} = \sqrt{(\gamma_{\rm cr} p_{\rm cr} + \gamma_{\rm th} p_{\rm th})/\rho }$ to estimate the CFL time step (\citealt{Courant1928}). The CFL time step is defined as 
\begin{eqnarray}
\Delta t_{\rm CFL} = {\rm (CFL\, number)}\,\times \,{\rm MIN}\left(\frac{\Delta x_{\rm i} }{u_{\rm max}^{\rm i}}\right)
\end{eqnarray}
where $\Delta x_{\rm i}$ is the separation between $i^{\rm th}$ and $(i+1)^{\rm th}$ grids, $u^{\rm i}_{\rm max}={\rm MAX}(|S_{\rm L}|, |S_{\rm R}|)$, $S_{\rm L}={\rm MIN} [(v^{\rm i-1}-c_{\rm s, eff}^{\rm i-1}), (v^{\rm i}-c_{\rm s, eff}^{\rm i})]$ and $S_{\rm R}={\rm MAX} [(v^{\rm i-1}+c_{\rm s, eff}^{\rm i-1}), (v^{\rm i}+c_{\rm s, eff}^{\rm i})]$ is the maximum wave speed between the left and right moving waves from the interface. {\textsc {TFH}} has gone through various test problems, and the comparisons with the publicly available code PLUTO (\citealt{Mignone2007}) are shown in Appendix \ref{app:codecheck}. 

For the runs in this paper, we set the solver to TS (see Appendix \ref{app:solver}) and the CFL number to $0.2$.  The details of important runs are given in Table \ref{tab:sim1}. In the following sections, we discuss the  terms on the RHS of equations (\ref{eq:mass}) - (\ref{eq:energycr}).
 \subsection{Ambient medium}
We consider a uniform ambient density and temperature. The ambient temperature is chosen such that the thermal pressure is $ 1$ eV cm$^{-3}$. For the runs with CRs, we set the CR pressure to $1$ eV cm$^{-3}$ (i.e., assuming equipartition between CRs and thermal gas).  For all cases, the initial velocity of gas is set to zero. We assume the metallicity of the injected materials to be solar, $Z=\,Z_{\rm \odot}$, the same as in the ambient gas, where $Z_{\rm \odot}$ is the solar metallicity.  
\subsection{Mass and energy deposition}
\label{subsec:injgen}
The source terms  $S_{\rho}$, $S_{e}/S_{\rm cr}$  represent the deposited mass and thermal/CR energy per unit time per unit volume. We add mass and energy in a region of radius $R_{\rm src}=1$ pc. The radius ($R_{\rm src}$) is chosen such that the energy loss rate is much less than the energy injection rate. This condition ensures that cooling at the initial stage will not suppress the production of strong shocks (for details, see section 4.2.1 in \citealt{Sharma2014}). 

The mass loss rate ($\dot{M}$) and the mechanical (wind) power of the source ($L_{\rm w}$) are related to the star formation rate (SFR) in the cloud and life-time of the individual stars. If $M_{\rm st}$ is the stellar mass in the cloud and $N_{\rm OB}$ is the number of OB stars with the mean-sequence lifetime $\tau_{\rm OB}$, then the ${\rm SFR}$ can be assumed to be $M_{\rm st}/\tau_{\rm OB}$ (\citealt{MacLow1988}). Assuming $M_{\rm st}\approx 50\,N_{\rm OB}\,M{\rm_{\odot}}$ and $\tau_{\rm OB}\approx 30\,{\rm Myr}$, we get ${\rm SFR}\simeq 1.67\times 10^{-6}\, N_{\rm OB}\,M{\rm_{\odot}\, yr^{-1}}$. We have also assumed that $\dot{M}=0.3\,{\rm SFR}$ and $10^{51}$ erg is the energy released in each supernova which yield
\begin{eqnarray}
\label{eq:Mdotsim}
\dot{M} \simeq 5\times 10^{-4}  \left(\frac{N_{\rm OB}}{10^3}\right) \left(\frac{\tau_{\rm OB}}{30\,{\rm Myr}}\right)^{-1}\ M{\rm_{\odot}\, yr^{-1}}
\end{eqnarray}
\begin{eqnarray}
\label{eq:Edotsim}
L_{\rm w}\simeq 10^{39} \left(\frac{N_{\rm OB}}{10^3}\right) \left(\frac{E_{\rm SN}}{10^{51}\,{\rm erg}}\right)\left(\frac{\tau_{\rm OB}}{30\,{\rm Myr}}\right)^{-1}\ {\rm erg\,s^{-1}}
\end{eqnarray}
If one assumes that the stellar winds and/or SNe deposit energy and mass continuously then the equations (\ref{eq:Mdotsim}) and (\ref{eq:Edotsim}) can be considered in the terms of SNe. For such case, the time interval between consecutive SNe is equivalent to $\delta t_{\rm SN} \simeq \tau_{\rm OB}/N_{\rm OB} = 0.03\, (N_{\rm OB}/10^{3})^{-1}$ Myr, and the total energy and mass deposited during that duration are $10^{51}$ erg and $15$ $M{\rm_{\odot}}$ respectively. Our choice gives the wind velocity $v_{\rm}\simeq\sqrt(2L_{\rm w}/\dot{M})\approx 2500$ km s$^{-1}$ which is consistent with Starburst99 (\citealt{Leitherer1999}) for a constant SFR or for a coeval star cluster.

To inject cosmic rays in the source region (c.f., section \ref{subsec:crinj}), we use a injection parameter $\epsilon_{\rm cr}$ (see equation (\ref{eq:ecrin})) to specify the fraction of the total input energy that given to CR (\citealt{Salem2013}; \citealt{Booth2013}). Therefore, $S_{\rm \rho}=\dot{M}/V_{\rm src}$, $S_{\rm th}=(1-\epsilon_{\rm cr}) L_{\rm w}/V_{\rm src}$ and $S_{\rm cr}=\epsilon_{\rm cr} L_{\rm w}/V_{\rm src}$. For fiducial runs, whenever CR injection at the source is mentioned, we set $\epsilon_{\rm cr} = 0.2$ otherwise it is set to zero. We explore the dependence of the results on various parameters in section \ref{sec:discussion}.
\subsection{Cooling \& Heating}
{\textsc {TFH}} uses the operator splitting method to include radiative cooling and heating. We use standard definition of cooling function as given below
\begin{eqnarray}
q^{-}_{\rm eff} = -\, n_{\rm i}\,n_{\rm e}\, \Lambda_{\rm N} + {\rm Heating}
\end{eqnarray}
where $\Lambda_{\rm N}$ is the normalized cooling function (\textsc{CLOUDY}, \citealt{Ferland1998}) and $n_{\rm i}/n_{\rm e}$ are the electron/ion number density. The region where the cooling energy loss becomes comparable to its thermal energy, {\textsc {TFH}} subcycles cooling. The number of sub-steps depends on the stiffness of cooling. We artificially stop cooling when the gas temperature goes below $10^{4}$ K. This corresponds to the shell temperature which is maintained due to the photo-heating by the radiation field of the driving source (for details see Figure 4 in \citealt{Gupta2016}). 
\subsection{Diffusion terms}
Thermal conduction and CR diffusion are also treated using operator splitting. Note that, for stability, the diffusion terms can have a much smaller time step compared to the CFL time step. {\textsc {TFH}} handles this by performing sub-cycling at each CFL time step.
\subsubsection{Thermal conduction}
\label{subsec:tc}
The thermal conduction flux $F_{\rm tc}$ is defined as:
\begin{eqnarray}
\label{eq:Ftc}
\vec{F}_{\rm tc} = \chi\, \vec{F}_{\rm classical}
\end{eqnarray}
where  $\vec{F}_{\rm classical}= -\kappa_{\rm th}\,\vec{\nabla} T$ and  $\kappa_{\rm th}= C\,T^{5/2}$ is the coefficient of thermal conduction (\citealt{Spitzer1962}) and $C$ is chosen to be $6\times 10^{-7}$ in cgs unit. The factor $\chi$ in equation (\ref{eq:Ftc}) limits the conduction flux when it approaches $\vec{F}_{\rm staturated}\approx(2\,k_{\rm  B}\,T/\pi m_{\rm e})^{1/2}n_{\rm e}k_{\rm  B}T_{\rm  e}$ (see \citealt{Cowie1977}) which is defined as $\chi = F_{\rm staturated}/(F_{\rm classical}+F_{\rm staturated})$.

The thermal conduction time step ($\Delta t_{\rm tc}$) is chosen to be
\begin{eqnarray}
\Delta t_{\rm tc} = ^{\rm MIN}_{\,\, \ \rm i}\left[ {\frac{\Delta x_{\rm i}^2}{2\, \kappa_{\rm th, avg}}}\right]
\end{eqnarray}
where $\kappa_{\rm th, avg}=  (\kappa_{\rm th, i-1}+\kappa_{\rm th,  i})/2$ is the cell averaged thermal conductivity.
\subsubsection{CR diffusion} 
CR diffusion follows a similar method as thermal conduction (section \ref{subsec:tc}) except that the cosmic ray diffusion flux $F_{\rm crdiff}$ is defined as:
\begin{eqnarray}
\label{eq:Ftc}
\vec{F}_{\rm crdiff} = -\kappa_{\rm cr}\vec{\nabla} \epsilon_{\rm cr}
\end{eqnarray}
where  $\kappa_{\rm cr}$ is the cosmic ray diffusion constant. 
Here we are only concerned with the hydrodynamical effects of CRs, but one should remember that $\kappa_{\rm cr}$ is an integral over 
the CR energy distribution function (equation (7) in \citealt{Drury1981}).
We consider  $\kappa_{\rm cr}=5\times 10^{26}\,{\rm cm^2\, s^{-1}}$ as a fiducial value which is consistent with the recent findings in the star forming/SNe region (e.g., \citealt{Ormes1988}; \citealt{Gabici2010}; \citealt{Li2010}; \citealt{Giuliani2010}) but smaller than the value usually adopted in the global ISM. We discuss the dependence of simulation results on this choice in section \ref{sec:discussion}.
\section{Results}
\label{sec:result}
\subsection{Ideal one-fluid ISB}
\label{subsec:idealone}
\begin{figure}
\centering
\includegraphics[height=2.25in,width=2.9in]{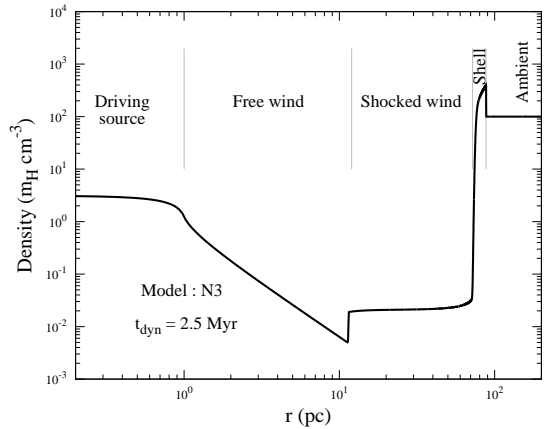}
\caption{Density profile of a standard one-fluid ISB. The horizontal axis represents the distance from the driving source ($r$ in pc) and vertical axis displays the density. The snap shot is taken at $t_{\rm dyn}=2.5$ Myr. For model details, see Table \ref{tab:sim1}.}
\label{fig:idealstruc}
\end{figure} 
To begin with, we recall the structure of a standard ISB (\citealt{Weaver1977}). We run an ideal one-fluid model by turning-off all microphysics (Table \ref{tab:sim1}). Figure \ref{fig:idealstruc} displays the density profile at $t_{\rm dyn}= 2.5$ Myr from this run. 

Starting from the left,  Figure \ref{fig:idealstruc} shows that the ISB consists of four distinct regions - (1) the source region where mass and energy are deposited, (2) the free-wind region where the wind expands adiabatically, (3) the shocked-wind region and (4) the swept-up ambient medium shell. Between the regions (3) and (4), there is a contact discontinuity (CD) which separates the ambient material from the ejecta material. 

In following sections, we include various microphysical processes one by one, for a better understanding of each process separately.
\subsection{CR injection in different regions}
\label{subsec:crinj}

\begin{figure*}
\centering
\includegraphics[height= 6.5in,width=6.2in]{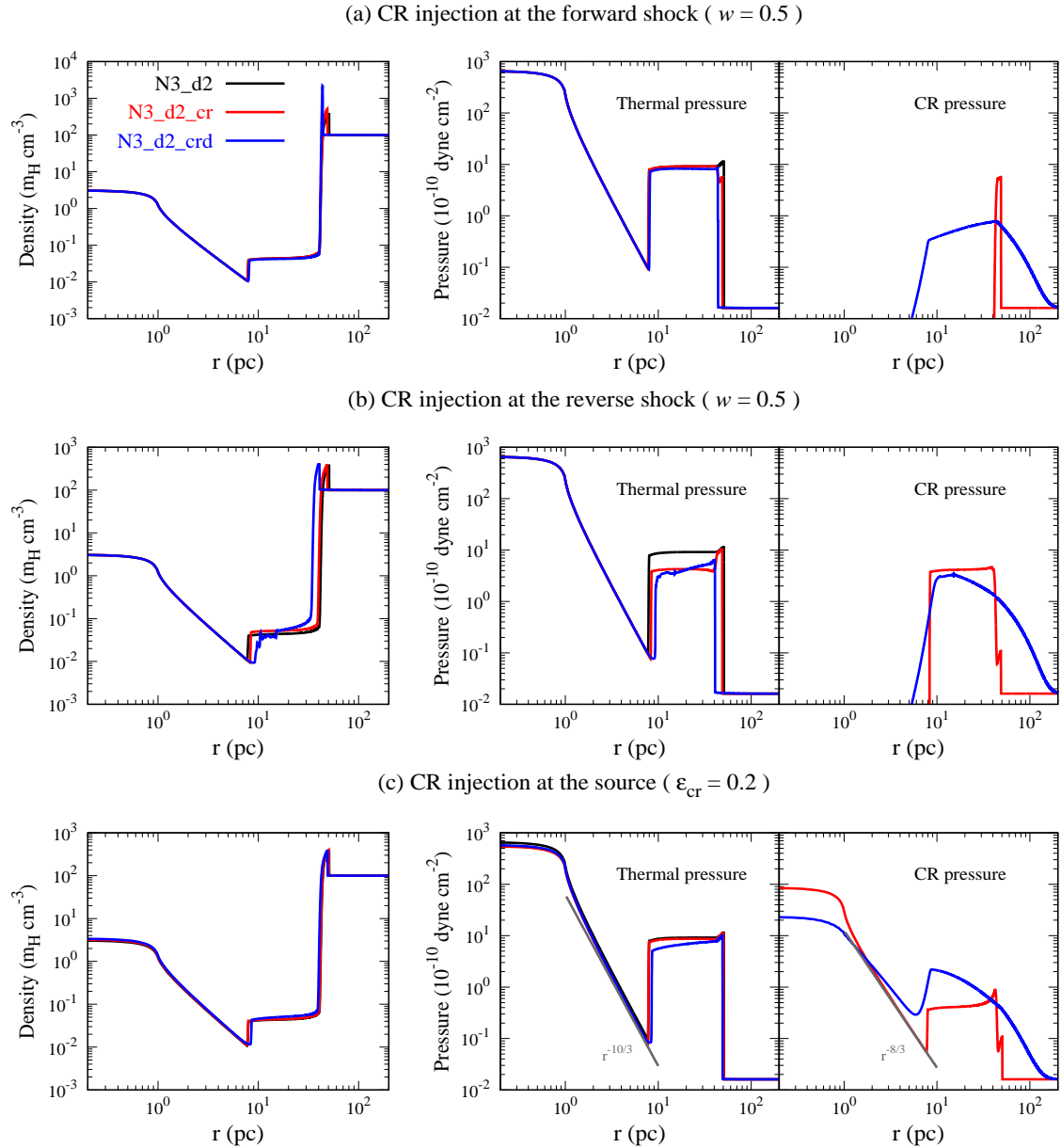}
\caption{CR injection in different regions. Each row displays the density and thermal/CR pressure profiles for a specific injection model (section \ref{subsec:crinj}) at $t_{\rm dyn}=1$ Myr. The black curves in all panels represent a one-fluid bubble. The red and blue both curves stand for two-fluid bubble, but CR diffusion ($\kappa_{\rm cr}=5\times 10^{26}\,{\rm cm^2\,s^{-1}}$) is on only for the blue curve. The grey lines in panel (c) display the free wind thermal pressure ($p_{\rm th}\propto r^{-10/3}$) and CR pressure ($p_{\rm cr}\propto r^{-8/3}$; equation (\ref{eq:pcrfree})) profiles respectively. This figure highlights the role of CR diffusion, and a self-consistent CR acceleration at the reverse shock (bottom row).}
\label{fig:crinj}
\end{figure*}

\subsubsection{At the forward/reverse shock}
\label{subsubsec:crinjshk}
In supernova remnant studies, the downstream CR pressure fraction is taken as $p_{\rm cr}=w\, p_{\rm tot}$ (see equation (\ref{eq:pcrin})) where the physical origin of $w$ is poorly understood (\citealt{Bell2014}). For the purpose of illustration, we choose $w=1/2$. For implementing CR pressure fraction, we have written a shock detection module which identifies the shock location and fixes the CR pressure accordingly. We have tested this module by comparing it with the blast wave self-similar solution of \citet{Chevalier1983} (see Appendix \ref{subappn:blast}). 

The results with CR injection at forward (top panel) and reverse (middle panel) shocks are displayed in Figure \ref{fig:crinj}. Black line stands for a standard one-fluid ISB (section \ref{subsec:idealone}). The red curve represents the case where only CR advection is taken into account. For the blue curve, in addition to advection, CR diffusion has been turned on.

With injection at forward shock, in the absence of CR diffusion (red curve), a fraction of the post-shock thermal pressure appears as CR pressure. In this case,  the contact discontinuity does not allow CR to be advected into the bubble i.e. the total pressure in the shell remains the same. Therefore, it changes the effective adiabatic index of the swept-up shell ($\gamma_{\rm eff}=[5+3\,w]/[3(1+w)]$), resulting in a higher density jump ($\simeq 5.5$) compared to the one-fluid case (for which the compression ratio is $4$). A more tangible difference appears when CR diffusion is turned on  (blue curves). In this case, CRs enter inside the bubble and also diffuse out of the shell. Therefore, the {\it total energy} in the shell is now reduced. This results in a much higher density jump, similar to the case with radiative cooling.

In the middle panel of Figure \ref{fig:crinj}, CRs have been injected at the reverse shock (RS). In the absence of CR diffusion (red curves), the CR pressure in the SW region remains almost the same as at the RS. As a result, the bubble size is slightly smaller than the standard bubble because of $\gamma_{\rm eff}=13/9$. In this case, one can use an effective $\gamma$ to determine the size of the bubble (see Table \ref{tab:const}). However, this conclusion is not valid if one considers CR diffusion. With CR diffusion the density jump at the RS is not sharp and the size of the bubble is smaller (compare the blue and black curves). 

The conclusions from this section are: (1) the CR diffusion plays an important role and (2) injection of CRs at the reverse shock can change the ISB structure.
\subsubsection{At the driving source region}
\label{subsubsec:crinjsrc} 
The numerical set-up for CR injection in driving source region is discussed in section \ref{subsec:injgen} and the results are shown in bottom panel of Figure \ref{fig:crinj}.

In absence of CR diffusion (red curves in panel (c) of Figure \ref{fig:crinj}), the free wind CR pressure profile follows $p_{\rm cr} \propto r^{-8/3}$ (see equation (\ref{eq:pcrfree})). The grey line in the bottom right most panel of Figure \ref{fig:crinj} displays this relation. The injected CR particles are advected up to the CD and the structure does not show any significant difference from that of the single fluid case. 

When diffusion is turned on (blue curves in panel (c)), we find that the CR pressure at the source region decreases which is expected because CRs diffuse from the high to low pressure region. The striking feature is that, at the RS, the CR pressure is quite large compared to the case of no diffusion. This jump in CR pressure is a property of two-fluid shocks as discussed in section \ref{subsubsec:insrc_ana}.

\begin{figure}
\centering
\includegraphics[height=2.25in,width=2.5in]{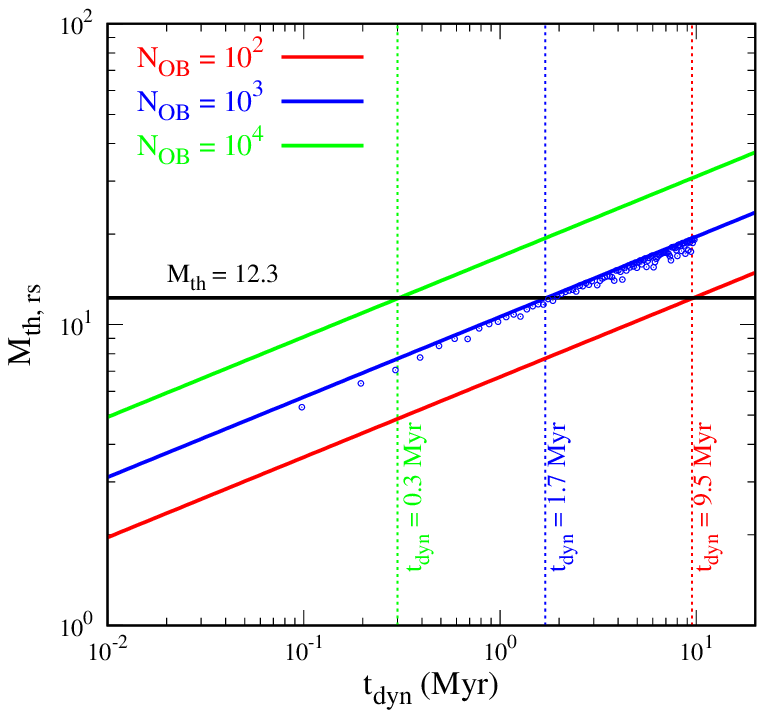}
\caption{Gas Mach number of the reverse shock as a function of dynamical time for three different $N_{\rm OB}$. The black line shows the critical Mach number for the globally smooth solution. The blue curve represents our fudicial model (also see equation (\ref{eq:Rrs_mach})) and the blue points show the data points from our simulation. This figure implies that, for large $N_{\rm OB}$, the globally smooth solution is achieved at early times.}
\label{fig:Mth_rs}
\end{figure}  

\begin{figure}
\centering
\includegraphics[height=2.25in,width=2.5in]{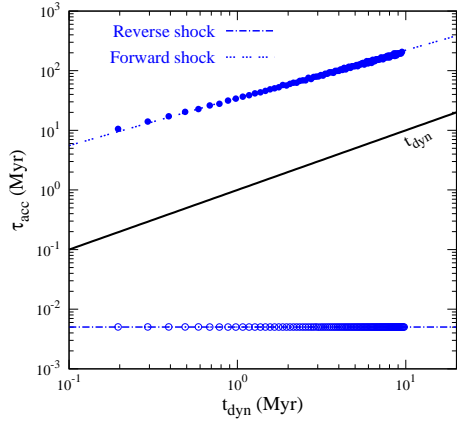}
\caption{The CRs acceleration time scale at the reverse shock and the forward shock as a function of dynamical time. The black line shows the dynamical time. The dashed and dashed-dotted blue lines are the analytic estimates from equation (\ref{eq:tau_acc}). The blue solid and empty circles show results from our simulation. Figure shows that, at the FS, $\tau_{\rm acc}$ is always greater than $t_{\rm dyn}$ explaining the reason for not seeing CR dominated smooth FS.}
\label{fig:tacc}
\end{figure}  

\begin{figure}
\centering
\includegraphics[height=2.9in,width=3.1in]{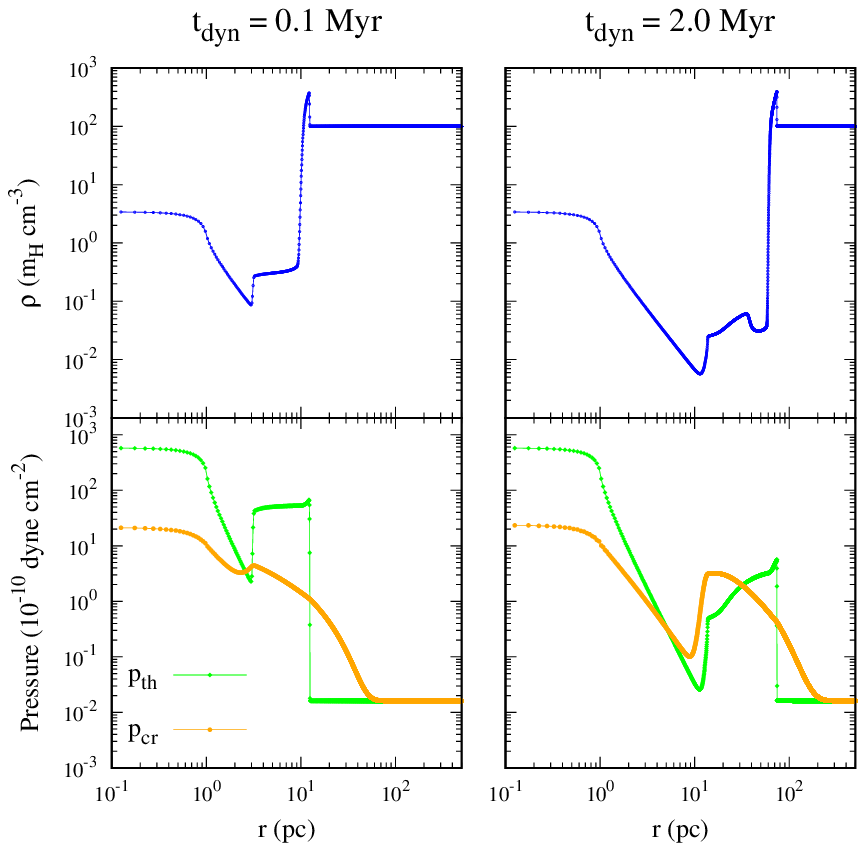}
\caption{Snapshot of the density and pressure profiles at $t_{\rm dyn}=0.1$ Myr (left panel) and $t_{\rm dyn}=2.0$ Myr (right panel). Top and bottom panels show the density and pressure profiles. This figure highlights the energy exchange between the thermal and non-thermal (CR) fluid at the reverse shock, and shows a smooth reverse shock at $t_{\rm dyn}=2{\,\rm Myr} > \tau_{\rm cri}=1.7$ Myr, consistent with equation (\ref{eq:tau_cri}).}
\label{fig:prs_tdyn}
\end{figure} 

Figure \ref{fig:Mth_rs} displays the time ($t_{\rm dyn}$) evolution of the reverse shock Mach number for three different values of $N_{\rm OB}$, where the blue colour stands for our fiducial choice. The solid lines in this figure display the analytical estimate of the RS gas Mach number (equation (\ref{eq:Rrs_mach})) and the blue circles represent the numerical results obtained from our fiducial run. This model shows that the RS satisfies the globally smooth condition after $t_{\rm dyn}\approx1.7$ Myr. The Mach number analysis for the FS shows that at early times, the FS also satisfies $M_{\rm th}>12.3$ (not displayed in this figure). However in our simulations, we do not see globally smooth shock at FS. The reason is discussed as follows.
 
For a large post shock CR pressure i.e. for efficient CR acceleration, the CR particles should cross the shock multiple times. \citet{Becker2001} found the critical Mach number for CR dominated post shock in steady state. Steady state assumes that CR particles have sufficient time to cross the shock multiple times. For time dependent calculation, we must consider CR acceleration time scale ($\tau_{\rm acc}$). \citet{Drury1983} discussed that if $\Delta t$ is the average time taken by the upstream CR particles to cross the shock and to return back from downstream (thereby, CRs complete one cycle), and $\Delta \mathcal{P}$ is the average momentum gain in one complete cycle, then 
\begin{eqnarray}
\label{eq:tau_acc}
\tau_{\rm acc}= \frac{\mathcal{P}_{\rm 1}}{\Delta \mathcal{P}/\Delta t}=\frac{ 3}{v_{\rm 1}-v_{\rm 2}}\left(\frac{\kappa_{\rm cr,1}}{v_{\rm 1}}+ \frac{\kappa_{\rm cr,2}}{v_{\rm 2}}\right) \, ,
\end{eqnarray} 
where subscripts $1,2$ stands for the upstream and down stream flow respectively (also see \citealt{Blasi2007}). Therefore, we can see the globally smooth solution only if $\tau_{\rm acc} \ll \tau_{\rm cri}<t_{\rm dyn}$.

Equation (\ref{eq:tau_acc}) can be written as $\tau_{\rm acc}\sim 6\,\kappa_{\rm cr}/v_{\rm 1}^{2}$ ($\equiv$ CR diffusion time scale). For the FS and RS, the condition $t_{\rm dyn}\gg\tau_{\rm acc}$ yields
\begin{eqnarray}
{\rm Forward\,shock:}\ t_{\rm dyn}\gg  11\ \kappa_{\rm 26}^{5}\, L_{\rm 39}^{-2}\,\rho_{\rm 2}^{2}\ {\rm Myr}\ {\rm ,}
\end{eqnarray}
\begin{eqnarray}
{\rm Reverse\,shock:}\ t_{\rm dyn}\gg  2\times 10^{-3}\ \kappa_{\rm 26}\, v_{\rm 3 }^{-2}\ {\rm Myr}  
\end{eqnarray}
where we have taken $v_{\rm 1}$ as the FS velocity in a uniform ambient medium ($\approx 0.6\,L_{\rm w}^{1/5}\,\rho^{-1/5}\,t_{\rm dyn}^{-2/5}$) and the wind velocity ($v_{\rm w}$) respectively. To illustrate this, we have estimated $\tau_{\rm acc}$ for the FS and RS (Model:${\rm N3\_d2}$; Table \ref{tab:sim1}), and the results are shown in Figure \ref{fig:tacc}.

Figure \ref{fig:tacc} plots $\tau_{\rm acc}$ as a function of dynamical time for FS and RS where CR diffusion constant is taken as $\kappa_{\rm cr}=5\times 10^{26}$ cm$^2$ s$^{-1}$. The dotted and dashed-dotted lines represent the analytic estimate of acceleration time (i.e., equation (\ref{eq:tau_acc})), and superposed on that, the blue solid and empty circles represent the simulation result for the respective shocks. This figure shows that $\tau_{\rm acc}$ is longer than $t_{\rm dyn}$ for the FS. This is so because the velocity of the upstream and down stream flow is very small (at least a factor of ten compared to RS) and decreases with time. Therefore a smooth solution is not achieved although it satisfies the upstream condition of the steady state model by \citet{Becker2001}. We conclude that for globally smooth solution, the condition is $\tau_{\rm acc}\ll \tau_{\rm cri}<t_{\rm dyn}$. This criterion limits the choice of $\kappa_{\rm cr}$ which is discussed in section \ref{subsec:crparameter}. Our two-fluid simulation results are consistent with this conclusion and the results are shown in Figure \ref{fig:prs_tdyn}.
 
Figure \ref{fig:prs_tdyn} shows the snapshots of density and thermal/CR pressure profiles at two different dynamical times. During the early evolution, the RS Mach number is less then $12.3$ (see Figure \ref{fig:Mth_rs}) and therefore the upstream CR does not dominate over the thermal pressure. As time evolves, after $\tau_{\rm cri}$, the gas Mach number exceeds the critical value and the shock becomes dominated by the CRs. This CR dominated shock is representative of globally smooth solution as first predicted by \citet{Drury1981}. In this case, the maximum CR pressure ($p_{\rm cr, max}$) in the SW region depends on the input source parameters, the ambient density and the dynamical time. We have found that, $p_{\rm cr, max}$ does not exceed the thermal pressure for the one-fluid ISB (i.e., $p_{\rm cr, max}\leq P$; see equation (\ref{eq:Pgen})), consistent with the total energy conservation.

The results discussed in this section do not include radiative cooling and thermal conduction i.e., all the changes are only due to CRs. We will discuss more realistic cases below.
\subsection{Toward a realistic model}
\label{subsec:realisticresult}
In the previous section, we have seen that the energy exchange between thermal and non-thermal particles becomes significant at the reverse shock. We have also noticed that when $p_{\rm cr}\gtrsim p_{\rm th}$, the cosmic ray fluid starts affecting the inner structures (see Figure \ref{fig:prs_tdyn}). For a radiative bubble, energy loss from the dense shell reduces its thermal energy. Therefore, we expect to see the impact of CRs at an earlier time than that in the adiabatic case. In order to get a more realistic picture, we discuss the effect of thermal conduction and then we turn on radiative cooling.
\subsubsection{Effect of thermal conduction}
\label{subsubsec:tceffect}
\begin{figure}
\centering
\includegraphics[height=2.25in,width=2.9in]{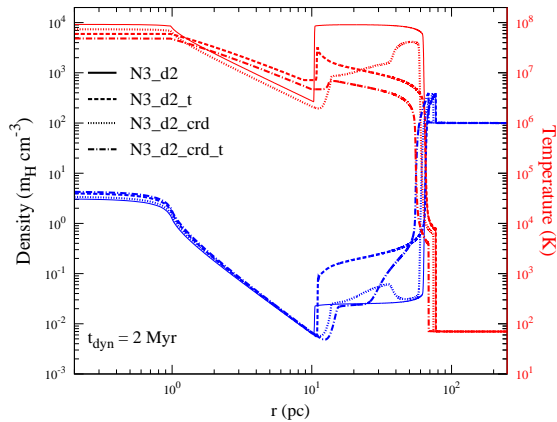}
\caption{ISB profile with and without thermal conduction. The blue and red curves show density and temperature profiles respectively. Each line style represents a specific model (see Table \ref{tab:sim1}). For two-fluid ISB models, CRs have been injected in the driving source and CR diffusion is on. Figure shows that CRs diminish the effect of thermal conduction.}
\label{fig:tceffect}
\end{figure}
To discuss the role of the thermal conduction in CR affected ISBs, we present a comparison of one-fluid and two-fluid models. Figure \ref{fig:tceffect} displays the density (blue curves) and temperature (red curves) profiles at $t_{\rm dyn}=2$ Myr. For a one-fluid ISB without thermal conduction (solid lines), the bubble temperature is high ($\sim 10^{8}$ K). With thermal conduction (dashed lines), the temperature drops to $\sim 10^{7}$ K, and the SW density increases. For a two-fluid ISB, the temperature is noticeably smaller than a one-fluid ISB {\it even without thermal conduction} (dotted lines). With thermal conduction it does not show a significant difference, except that, it smoothens the temperature near the CD (dash-dotted lines). Therefore, we conclude that CRs reduce the effect of thermal conduction. We have reached the same conclusion for ISBs with radiative cooling. In the following sections, we continue our discussion without thermal conduction.

\begin{figure}
\centering
\includegraphics[height=2.9in,width=2.9in]{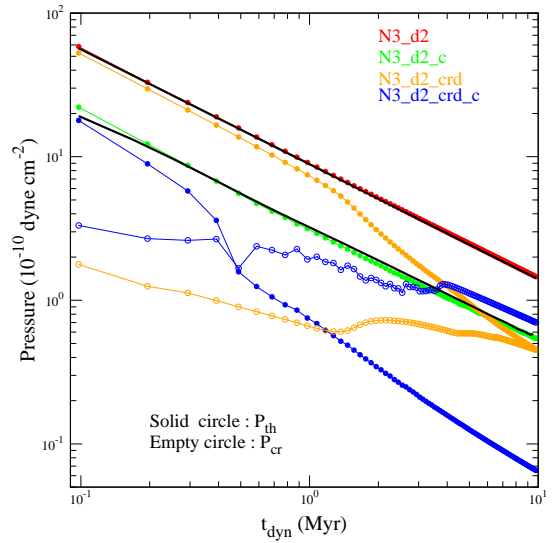}
\caption{Time evolution of volume averaged hot gas pressure. The solid and empty circles denote thermal pressure and CR pressure respectively. Different colours represent the results for different models (see Table \ref{tab:sim1}). Black and grey lines display the expected (equation (\ref{eq:prs_anafit})) thermal pressure for a one-fluid bubble. This figure shows that the thermal pressure does not follow equation (\ref{eq:prs_anafit}) for the CR affected ISB.}
\label{fig:estprs}
\end{figure}

\subsubsection{Volume averaged quantities}
Figure \ref{fig:estprs} displays the volume averaged hot gas ($>10^{5}$ K) pressure and the CR pressure as functions of time for various models. The solid and empty circles display the thermal and CR pressure respectively. The black solid lines (top curve : adiabatic and bottom curve : radiative) display equation (\ref{eq:prs_anafit}) ($\gamma = 5/3$, $s=0$) where the shell velocity is estimated from the simulation. In this case, the analytical result agrees with the simulation (as shown by the concurrence of red points with the top black line, and green points with the bottom black line). For two-fluid ISBs (yellow and blue curves), as time evolves, the thermal pressure deviates from this relation. This deviation is mainly because in a two-fluid ISB, a major fraction of the free wind kinetic energy goes to CR.
\subsection{Evolution of different energy components}
\label{subsubsec:energyevol}
\begin{figure}
\centering
\includegraphics[height=7in,width=2.9in]{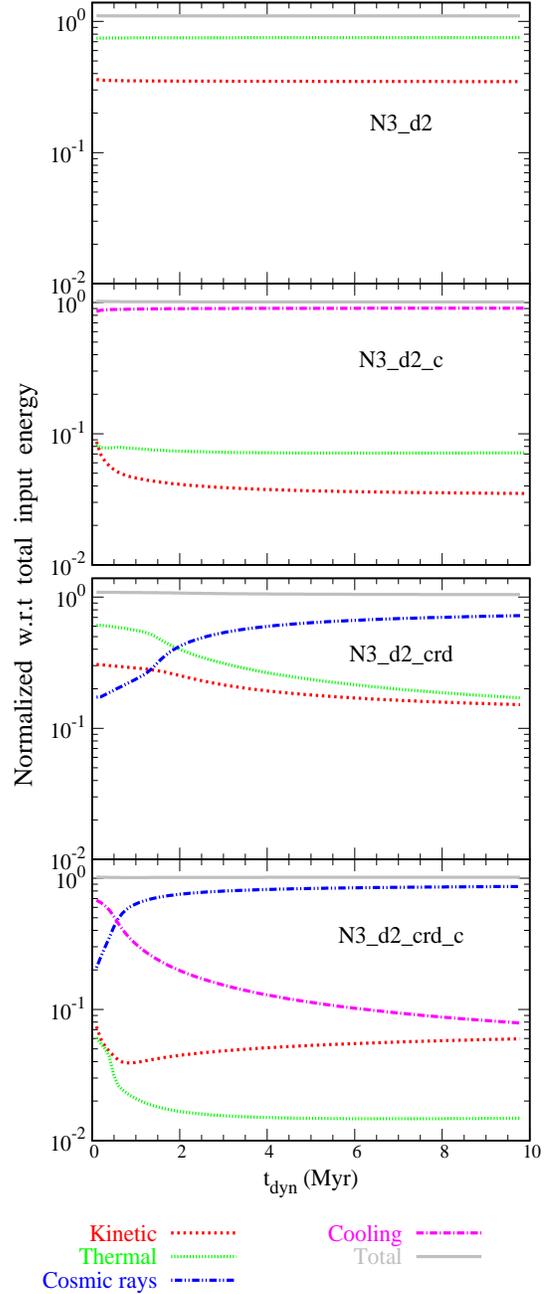}
\caption{Time evolution of energy fraction (normalized to total injected energy) in four different cases. The upper two panels represent the adiabatic and radiative one-fluid ISB. The lower two panels stand for the adiabatic and radiative two-fluid ISB. For the two-fluid case, the CRs has been injected in the source region ($\epsilon=0.2$). For all panels, red colours stand for kinetic energy, green for the thermal energy and blue for CR energy. The magenta colour represents radiative energy losses. This figure shows that for two-fluid realistic case (i.e., when all microphysics are truned on; Model : ${\rm N3\_d2\_crd\_c}$ in Table \ref{tab:sim1}), CRs dominate over other energy components.}
\label{fig:evolenergy}
\end{figure} 
Now we come to the evolution of the kinetic/thermal/CR energy and the radiative loss for four different models. For all cases, we estimate the change of the entity ${\rm x}$ in the simulation box and normalized it w.r.t to the total deposited energy by that epoch. These entities have been estimated using 
\[\Delta E_{\rm x}^{\rm t}  = \int_{\rm V} dV\, e_{\rm x}^{\rm t} - \int_{\rm V} dV\, e_{\rm x}^{0} \ \,\ ,\ \, E_{\rm IN}^{\rm t} = \int_{\rm t=0}^{\rm t} dt\, L_{\rm w} \  ,\]
\begin{equation}
\label{eq:change_energy}
E_{\rm loss}= \int_{\rm t=0}^{\rm t} dt\, \int_{\rm V} dV\, n_{\rm e}n_{\rm i}\Lambda_{\rm N}\ ,
\end{equation}
where ${\rm x}$ refers to the kinetic/thermal/CRs energy, $E_{\rm loss}$ is the radiative loss  and $ E_{\rm IN}^{\rm t}$ is the total deposited energy at the source. Our results are displayed in Figure \ref{fig:evolenergy}. 

The top panel of this figure displays the dynamical evolution of the kinetic energy (KE, red curve) and thermal energy (TE, green curve) for a one-fluid adiabatic ISB. The cooling is turned on in the second panel. In this case, the magenta curve represent the cooling losses which shows that almost $85-90\%$ energy is radiated from the ISB. Therefore, the total energy retained in the ISB is $10-15\%$. One should note that this fraction may change depending on the density and metallicity of ISM. The lower two panels show the results for a two-fluid ISB. The third panel shows the adiabatic bubble with CR diffusion and fourth panel shows radiative two-fluid bubble with CR diffusion. A comparison between second and bottom (magenta colour) panels demonstrate that CRs suppress cooling losses.

\section{Discussion}
\label{sec:discussion}
In this section we go beyond the fiducial models and explore the parameter space (sections \ref{subsec:crparameter}, \ref{subsec:ambientdepen}, \ref{subsec:inputenergy}). We show a comparison of one-fluid and two-fluid runs, in terms of the distribution of hot gas at different dynamical times (section \ref{subsec:dentemmap}). We also show the total energy gain by the CRs and discuss its dependence on various parameters (section \ref{subsec:energygain}). 
\begin{figure*}
\centering
\includegraphics[height=1.9in,width=6.7in]{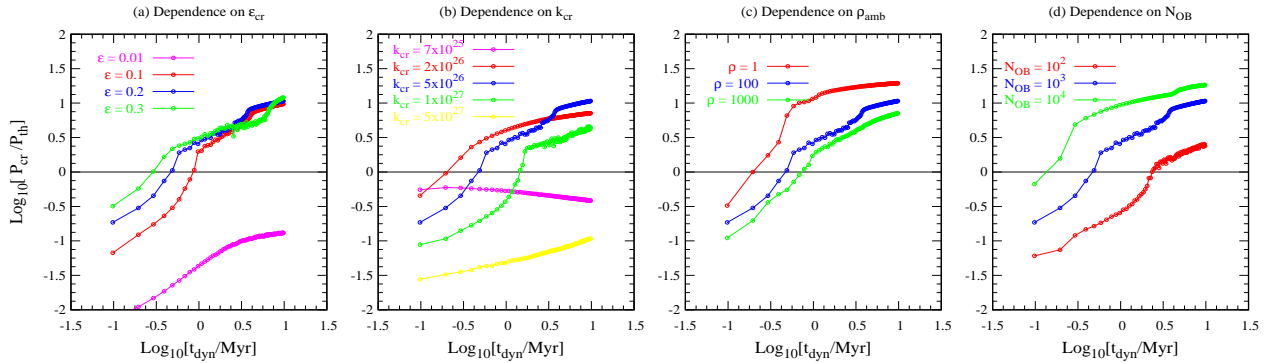}
\caption{Comparison between the volume averaged CR pressure and thermal pressure of the hot gas and its dependence on various parameters. The vertical and horizontal axes of all panels represent the ratio of $P_{\rm cr}$ to $P_{\rm th}$ and the dynamical time ($t_{\rm dyn}$) respectively. Any point above the black horizontal line denotes a CR dominated ISB. For all panels, the blue colour refers to the fiducial run. For all models, the radiative cooling and CR diffusion are switched on, and CRs are injected in the source region. For details, see sections \ref{subsec:crparameter}, \ref{subsec:ambientdepen} and \ref{subsec:inputenergy}.}
\label{fig:depend}
\end{figure*}
\subsection{Choice of CR parameters}
\label{subsec:crparameter}
To begin with, we explore the parameter space of CR injection parameter $\epsilon_{\rm cr}/w$ and diffusion constant ($\kappa_{\rm cr}$). The CR injection parameter $w$ (at shock, see equation (\ref{eq:pcrin})) or $\epsilon_{\rm cr}$ (at driving source, equation (\ref{eq:ecrin})) are crucial parameters of two-fluid ISBs, although their origin and values are not known (\citealt{Bell2014}). We treat them as a free parameters. As the dependence of the result on $w$ is easily predictable in the one-fluid model (see section \ref{subsec:crinj}), here, we present the result for different values of $\epsilon_{\rm cr}$ and $\kappa_{\rm cr}$. 

To visualize the dependence, we have estimated the ratio of volume averaged CR pressure to thermal pressure of the hot gas (temperature $>10^{5}$ K) and displayed them as a function of dynamical time in Figure \ref{fig:depend}. Any point above the black horizontal line represents a CR dominated ISB. Panel (a) shows that, for a larger value of $\epsilon_{\rm cr}$, the bubble becomes CR dominated at an early time. This is consistent because a larger $\epsilon_{\rm cr}$ increases the upstream CR pressure at the reverse shock. Panel (b) shows that, ISB becomes CR dominated if $10^{26}\lesssim \kappa_{\rm cr}/({\rm cm^{2}\, s^{-1}})\lesssim 3\times 10^{27}$. If $\kappa_{\rm cr}$ is below $< 10^{26}\,{\rm cm^{2}\, s^{-1}}$, then CR diffusion is almost negligible resulting in an unaffected ISB. Whereas, as $\kappa_{\rm cr}$ increases, CRs diffuse out of the ISB, and therefore, a one-fluid ISB model (discussed in section \ref{subsec:onefluidtheory}) is good enough to describe its structure.
\subsection{Dependence on the ambient density}
\label{subsec:ambientdepen}
The choice of the ambient density is an important parameter while comparing the theoretical result with observations. Most of the observations provide the density of the photo-ionized shell, but, beyond that, it is decipher to predict it from observations. Here we discuss the role of the ambient density for the two-fluid ISBs.

To scan the ambient density parameter space, we select two densities: $\rho =1 \,m{\rm_H\, cm^{-3}}$ and $\rho= 10^{3}\,m{\rm_H\, cm^{-3}}$ (recall that our fiducial density is $\rho= 10^{2}\,m{\rm_H\, cm^{-3}}$). The nature of the ISB can be inferred from equation (\ref{eq:tau_cri}), which states that for a high ambient density, the ISB takes a longer time to attain a globally smooth solution. Simulations agree with analytic estimates as shown in panel (c) of Figure \ref{fig:depend}. Figure shows that for a lower ambient density affects the interior at an earlier time. 
\subsection{Dependence on $N_{\rm OB}$}
\label{subsec:inputenergy}
The dependence on  $N_{\rm OB}$ is quite clear from equation (\ref{eq:tau_cri}). A larger number of $N_{\rm OB}$ corresponds to a higher wind luminosity and mass-loss rate ($\tau_{\rm cri}\propto \dot{M}^{-5/8}L_{\rm w}^{-1/8}\sim N_{\rm OB}^{-3/4}$). This means that the reverse shock can become smooth at early times. The simulation results shown in panel (d) of Figure \ref{fig:depend} are consistent with this. Therefore, superbubble reverse shocks, with high Mach numbers, are a promising site for CR acceleration.
\begin{figure*}
\centering
\includegraphics[height=2.9in,width=6.2in]{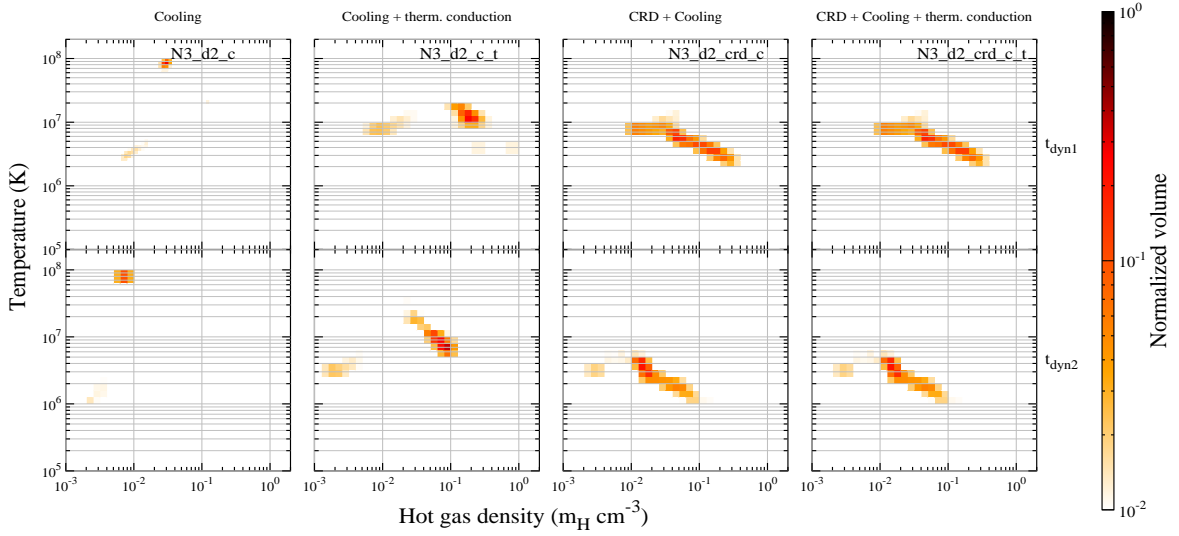}
\caption{Comparison of  hot gas properties for the one-fluid (first and second columns) and two-fluid (third and fourth column) ISBs in the absence (First and third column) or presence (second and fourth column) of thermal conduction. This figure display the snapshots of density-temperature distribution of the hot gas ($>10^{5}$ K) at two different dynamical times ($t_{\rm dyn1} =0.5$ Myr and $t_{\rm dyn1} =2.5$ Myr). Starting from the left, the first/second column represents a one-fluid ISB without/with thermal conduction, the third column displays the two-fluid model with CR diffusion (CRD) and the fourth column displays the two-fluid model with CR diffusion (CRD) plus thermal conduction. This figure shows that CR significantly affects the distribution of thermal gas inside the ISBs. We get similar pattern for $N_{\rm OB}=10^{2}$ and $N_{\rm OB}=10^{4}$.}
\label{fig:nxTx}
\end{figure*} 
\subsection{Density-temperature of hot gas}
\label{subsec:dentemmap}
Figure \ref{fig:nxTx} shows the density-temperature distribution of hot gas. To estimate this, first, we divide the data into different density channels from $\rho=10^{-3}$ to $2\,m{\rm_{H}\,cm^{-3}}$ $(\delta \rho=10^{0.1}\,m{\rm_H\, cm^{-3}})$. For each density channel, we create temperature bins ($\delta T=10^{0.1}\,{\rm K}$ and $10^{5}\leq T/{\rm K}\leq 2\times 10^{8}$), and then, we calculate the hot gas volume within a given density-temperature bin. The normalization is such that the same colour corresponds to an identical volume fraction in all panels.

For all panels of Figure \ref{fig:nxTx}, cooling is on. The left most panel displays a one-fluid ISB without thermal conduction. The temperature of the gas is $\sim 10^{8}$ K and density is low (as expected from equations (\ref{eq:Tsw}) and (\ref{eq:rhosw})). When thermal conduction is turned on (second panel from the left), the temperature drops to $10^{7}$ K. In this case, the mass density is high because of the mass evaporation from the dense shell (see section \ref{subsec:tc}). Most interesting processes take place when we switch to a two-fluid ISB (third and fourth panels; $\kappa_{\rm cr}=5\times 10^{26}\,{\rm cm^{2}\, s^{-1}}$). In this case, even without thermal conduction (third panel), the temperature drops to $(1-5)\times 10^{6}$ K. The thermal conduction (right most panel) does not show any significant difference because the lowered temperature diminishes the effect of thermal conduction. Figure shows that the CR affected ISBs can have temperature much lower than that of a one-fluid ISB.
\begin{figure*}
\centering
\includegraphics[height=1.9in,width=6.7in]{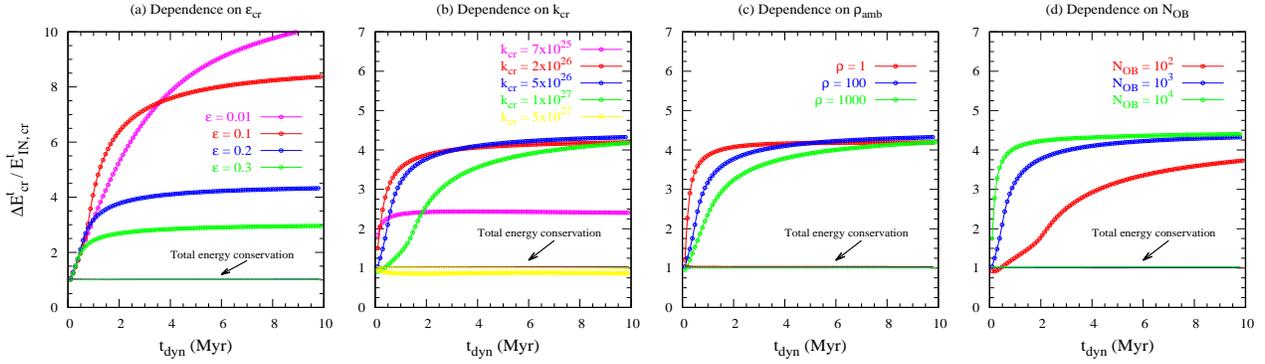}
\caption{Time evolution of the net change in CR energy (normalized w.r.t to the injected CR energy). For all panels, the blue curve refer to the fiducial run (for different resolutions, see Figure \ref{fig:resol2}). The horizontal lines near unity display total energy conservation in the respective runs.}
\label{fig:fracgaincr}
\end{figure*} 
\subsection{Total energy gain by CRs}
\label{subsec:energygain}
In previous sections, we have seen that for two-fluid ISBs, CR can gain energy from thermal particles. Here we show the energy gain by CRs and plot it as a function of dynamical time. To the best of our knowledge, this is the first time that the net gain of CR energy ($\Delta E^{\rm t}_{\rm cr}$) has been presented.

To obtain CR energy gain, we estimate $\Delta E^{\rm t}_{\rm cr}$ by following the same method as discussed in section \ref{subsubsec:energyevol}. The fraction of energy gain by the CRs $\Delta E^{\rm t}_{\rm cr}/E^{\rm t}_{\rm IN, cr}$ (where, $E^{\rm t}_{\rm IN, cr}=\epsilon_{\rm cr}E^{\rm t}_{\rm IN}$, see equation (\ref{eq:change_energy})) and its dependence on various parameters are displayed in Figure \ref{fig:fracgaincr}. To prevent CR overflow from the computational box (especially for the low density ambient, large $\kappa_{\rm cr}$ and $\epsilon_{\rm cr}$), we set $\rm r_{max} = 1400.1$ and choose number of grid points $n=28000$ (Table \ref{tab:sim1}). We have tested the total energy efficiency of the simulation box which is defined as 
\begin{equation}
\label{eq:simenergyefficiency}
 \rm{Energy\,efficiency\,of\,box} =\frac{\sum_{\rm x=th, ke, cr}\,\Delta E^{\rm t}_{\rm x} + E^{\rm t}_{\rm loss}}{E^{\rm t}_{\rm IN}}
 \end{equation}
where ${\rm x}$ is the kinetic/thermal/CRs energy and $E_{\rm loss}$ is the radiative energy loss (also see equation (\ref{eq:change_energy})). All the runs fulfill the energy conservation with an accuracy $>97\%$, shown by the horizontal lines close to unity in Figure \ref{fig:fracgaincr}.

The panel (a) of Figure \ref{fig:fracgaincr} shows that the CR energy gain fraction increases as $\epsilon_{\rm cr}$ decreases. This is reasonable because at the shock the upstream kinetic energy gets converted to the downstream CR energy via CR diffusion (although the energy transfer is insufficient for it to become a CR dominated ISB, see panel (a) in Figure \ref{fig:depend}). Panel (b) shows that CRs gain energy if $10^{26}\lesssim \kappa_{\rm cr}/({\rm cm^{2}\, s^{-1}})\lesssim 3\times10^{27}$, consistent with the conclusion of section \ref{subsec:crparameter}. Panel (c) shows that, an ISB expanding in a low density ISM achieves maximum energy at an early time. Panel (d) confirms that a large $N_{\rm OB}$ can be a promising source of CRs.
\section{Astrophysical implications}
\label{sec:anaobs}
\begin{table}
\caption{The observed ISB parameters.}
\label{tab:obs_bubbles}
  \raggedleft
 \begin{tabular}{l} 
 \includegraphics[height=3.5in,width=3.35in]{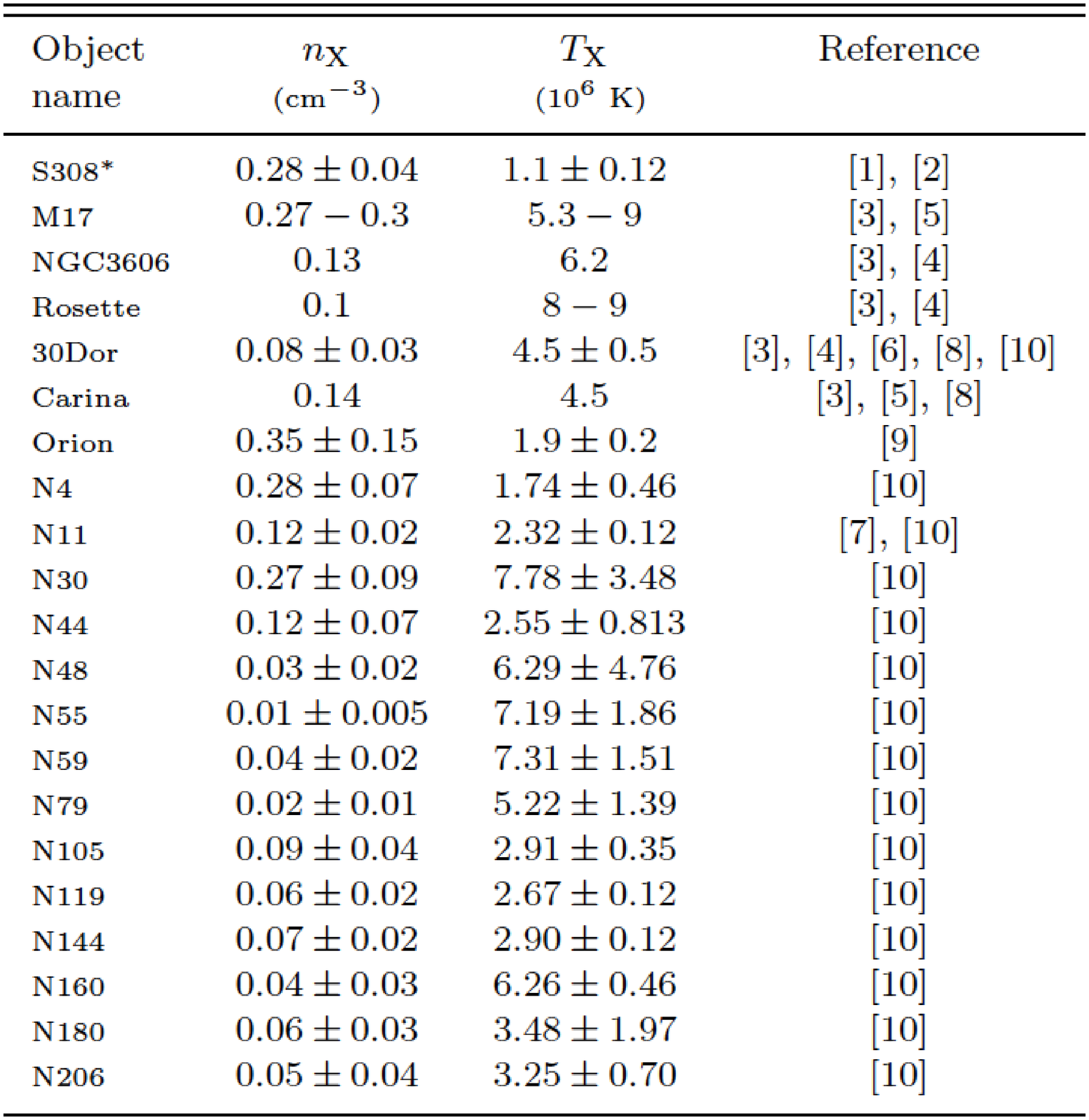}
 \end{tabular}\\
 \raggedright{[1] \citealt{Chu2003a}, [2] \citealt{Toala2013}, [3] \citealt{Rosen2014} [4] \citealt{Townsley2003}, [5] \citealt{Dunne2003}, [6] \citealt{Pellegrini2011}, [7]  \citealt{Maddox2009}, [8] \citealt{Lopez2011}; [9] \citealt{Gudel2008}, [10]\citealt{Lopez2014}. ($^*$ ISB containing two stars).}
\end{table}

X-ray diagnostic is an important tool to study the interior of ISBs. Since the bremsstrahlung emissivity decreases exponentially above the gas temperature, one can find the temperature of the plasma from the X-ray spectrum. This method is very useful to determine the best-fit temperature of the X-ray emitting gas (\citealt{Dunne2003}; \citealt{Townsley2006}; \citealt{Lopez2011}). The shape of spectra also tells whether emissions are coming from the thermal particles or from non-thermal particles (\citealt{Dunne2003}). The non-thermal emission can be used for modeling CR energy (\citealt{Helder2009}).

For most of the observed ISBs, the swept-up shell is quite clumpy which allows the X-ray radiation to come out, and helps one to estimate the hot gas density. The density of the hot gas is mostly obtained from the emission measure.  This technique requires assuming a hot gas volume filling factor, which is taken to be $1/2-1$. However, we should note that this may introduce an error in the hot gas density.

\begin{figure}
\centering
\includegraphics[height=2.25in,width=2.4in]{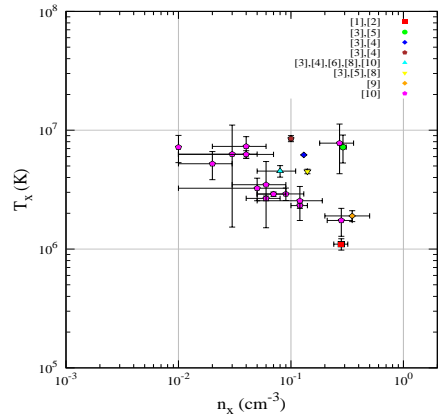}
\caption{The best-fit number density ($n_{\rm x}$) and temperature ($T_{\rm x}$) of the X-ray emitting plasma in the observed ISBs. All the plotted data points are taken from the  literature (details given in Table \ref{tab:obs_bubbles}). This figure shows that, for none of the ISBs, the X-ray temperature exceeds $10^7$ K ($\equiv 0.86$ keV).}
\label{fig:obs_isb}
\end{figure} 

The best fit temperature ($T_{\rm x}$) and the density ($n_{\rm x}$) of the X-ray emitting hot gas in the observed ISBs are given in Table \ref{tab:obs_bubbles}. All these data points are taken from published results (references  given in the caption of Table \ref{tab:obs_bubbles}). Figure \ref{fig:obs_isb} visualizes the content of this table. For all these ISBs, the dynamical age is less then $10$ Myr (\citealt{Lopez2014}). This figure shows that the number density of X-ray gas is $n_{\rm x}=(0.01\hbox{--}0.3)\,{\rm cm^{-3}}$ and the temperature is $\approx (1\hbox{--}9) \times 10^6$ K. 

If one considers that thermal conduction is the only physical process controlling the interior temperature, then the one-fluid model can be used to explain some of the observed data points. However, as we have mentioned in section \ref{sec:analytical}, thermal conduction can be affected by magnetic fields. Alternatively, CRs can be responsible for lowering the temperature of the hot gas which diminishes the mixing between the ambient and ejecta material via evaporation. Therefore, if an observation finds that the temperature of the hot gas does not agree with the wind velocity (see equation (\ref{eq:Tsw})) and the hot gas material is different from the ambient material (usually done by determining metallicity), then that would be a promising evidence for a CR affected ISB. 

A comparison of simulation results with observations shows that the temperature matches well if CR is injected at the driving source region with a CR energy fraction $\epsilon_{\rm cr}\sim 10-20\%$ of total input energy and a CR diffusion constant $\kappa_{\rm cr}\approx (0.1-3)\times 10^{27}\, {\rm cm^2\,s^{-1}}$. A higher value of CR diffusion constant decouples the CR energy and the thermal energy resulting in an unmodified superbubble. Therefore, if the diffusion constant is very high $(\gg10^{27} \, {\rm cm^2\,s^{-1}})$, then, regardless of whether or not ISBs are sites of CR acceleration, the thermal X-ray temperature will not be a good diagnostic of the presence of CR.

It is worth mentioning that a one-dimensional hydrodynamic analysis can not capture magnetized thermal conduction and cosmic ray diffusion. These need further investigation have been left for future work.
\section{Summary}
\label{sec:summary}
In this paper we have presented a two-fluid model of the interstellar bubbles by considering CR as a second fluid. Our work can be seen as a generalization of two standard theories of outflows (1) CR affected blast wave (first modeled by \citealt{Chevalier1983}) and (2) one-fluid interstellar bubble (\citealt{Weaver1977}). The main results from this work are given below:
\begin{enumerate}
\item {\it Dynamics without CRs}: We have found that the thermal pressure inside the bubble follows a robust relation which holds for an arbitrary density profile (equation (\ref{eq:rhof})) even with radiative cooling and thermal conduction. According to this, the volume averaged pressure inside an ISB is $\approx \rho\, v_{\rm sh}^2$ (see equation (\ref{eq:prs_anafit})), where $v_{\rm sh}$ is the velocity of the expanding shell and $\rho$ is the ambient density. Therefore, the deviation from this relation can be considered as an indication of the presence of CRs (see Figure \ref{fig:estprs}).
\item {\it CR injection and its effect}: The effect of CRs depends on (1) where the seed relativistic particles are injected and (2) what fraction of the total injected energy/post shock pressure goes into it. Since an ISB consists two shocks, one can inject CRs at the forward and/or at the reverse shock. The injection at the forward shock does not change the interior structure (panel (a) in Figure \ref{fig:crinj}). This makes ISBs different from a two-fluid SNe shock because the center of the blast wave is dominated by CRs if the injection is done at the shock (\citealt{Chevalier1983}; also see Figure $1$ in \citealt{Bell2014}). The injection at the reverse shock can reduce the thermal pressure inside the ISB. The CR injection at shocks is described by an ad-hoc parameter (denoted by $w$, definition is given in section \ref{subsec:crinj}) which does not capture the actual physics of two-fluid shock. However, a self-consistent evolution is obtained when CRs are injected in the driving source region (panel (c) in Figure \ref{fig:crinj}). In this case, depending on the CR diffusion constant, the reverse shock can show all possible solutions of a two-fluid shock predicted by \citealt{Drury1981} (Figure \ref{fig:prs_tdyn}).
\item {\it The importance of CR diffusion constant} : The key parameter in two-fluid model is the CR diffusion constant ($\kappa_{\rm cr}$). One can see a significant difference between the ISBs with and without CRs only if $\tau_{\rm acc}\ll \tau_{\rm cri} < t_{\rm dyn}$ (section \ref{subsubsec:crinjsrc}) where $\tau_{\rm acc}$ is the CR acceleration time scale (equation (\ref{eq:tau_acc}); analogous to diffusion timescale), $\tau_{\rm cri}$ (equation (\ref{eq:tau_cri})) is the time taken by the reverse shock to exceed the the critical Mach number (\citealt{Becker2001}) for a globally smooth reverse shock profile and $t_{\rm dyn}$ is the dynamical time. We have found that this condition is fulfilled if $10^{26}\lesssim \kappa_{\rm cr}/({\rm cm^{2}\, s^{-1}})\lesssim 3\times 10^{27}$ and in the case of ISBs with a large number OB stars ($N_{\rm OB}\geq 10^{2}$) (panels (b) and (d) in Figures \ref{fig:depend} and \ref{fig:fracgaincr}). This supports the suggestion in the literature from phenomenogical studies, that massive compact stellar associations can be promising source of CRs (\citealt{Higdon2006}; \citealt{Ferrand2010}; \citealt{Lingenfelter2012}).
\item {\it Observational signatures}: An indirect evidence for the presence of CRs in ISBs can be inferred from the temperature of the X-ray emitting plasma. We have found that the CR affected bubble can have temperature $1-5\times 10^{6}$ K even in the absence of thermal conduction (Figure \ref{fig:nxTx}) which can explain the X-ray temperature in the observed ISBs. 
\end{enumerate}

The model presented in this paper is admittedly idealized, which can be extended to a more realistic scenario. $3$D MHD simulations of two-fluid ISBs will be important to shed further light on the question of CR origin. 
\section*{Acknowledgements}
DE acknowledges support from an ISF-UGC grant, the Israel-US Binational Science Foundation, and the Joan and Robert Arnow Chair of Theoretical Astrophysics. PS acknowledges an India-Israel joint research grant (6-10/2014[IC]). SG thanks SPM fellowship of CSIR, India for financial support.

\footnotesize{

\appendix
\section{Code check}
\label{app:codecheck}
We have performed several standard test problems to check our code TFH. Here we present two of them: (1) the shock tube problem (\citealt{Sod1978}) in cartesian geometry (section \ref{subapp:1dshk}) and (2) the blast wave problem (\citealt{Sedov1946}; \citealt{Taylor1950}) in spherical geometry (section \ref{subappn:blast}). For both cases, we present one-fluid and two-fluid solutions. We also present a test problem for the diffusion module (thermal conduction/CR diffusion) in section \ref{subappn:diffusion}. We have compare our results with analytical solutions and also with the publicly available one fluid code PLUTO (\citealt{Mignone2007}).
\begin{figure*}
\begin{minipage}[b]{0.45\linewidth}
\centering
\includegraphics[height=2.8in,width=2.9in]{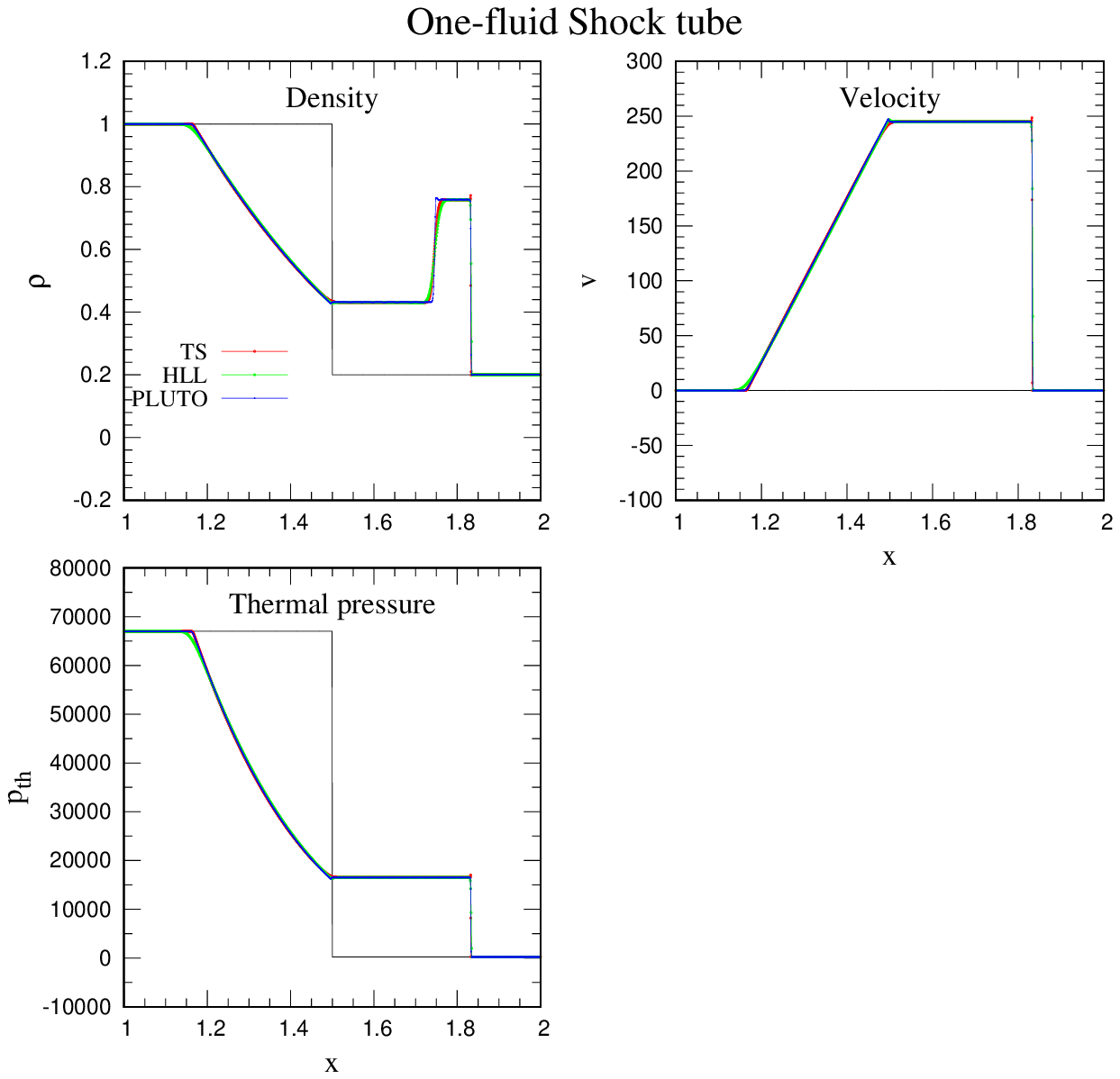}
\end{minipage}
\hspace{0.05cm}
\begin{minipage}[b]{0.45\linewidth}
\centering
\includegraphics[height=2.8in,width=2.9in]{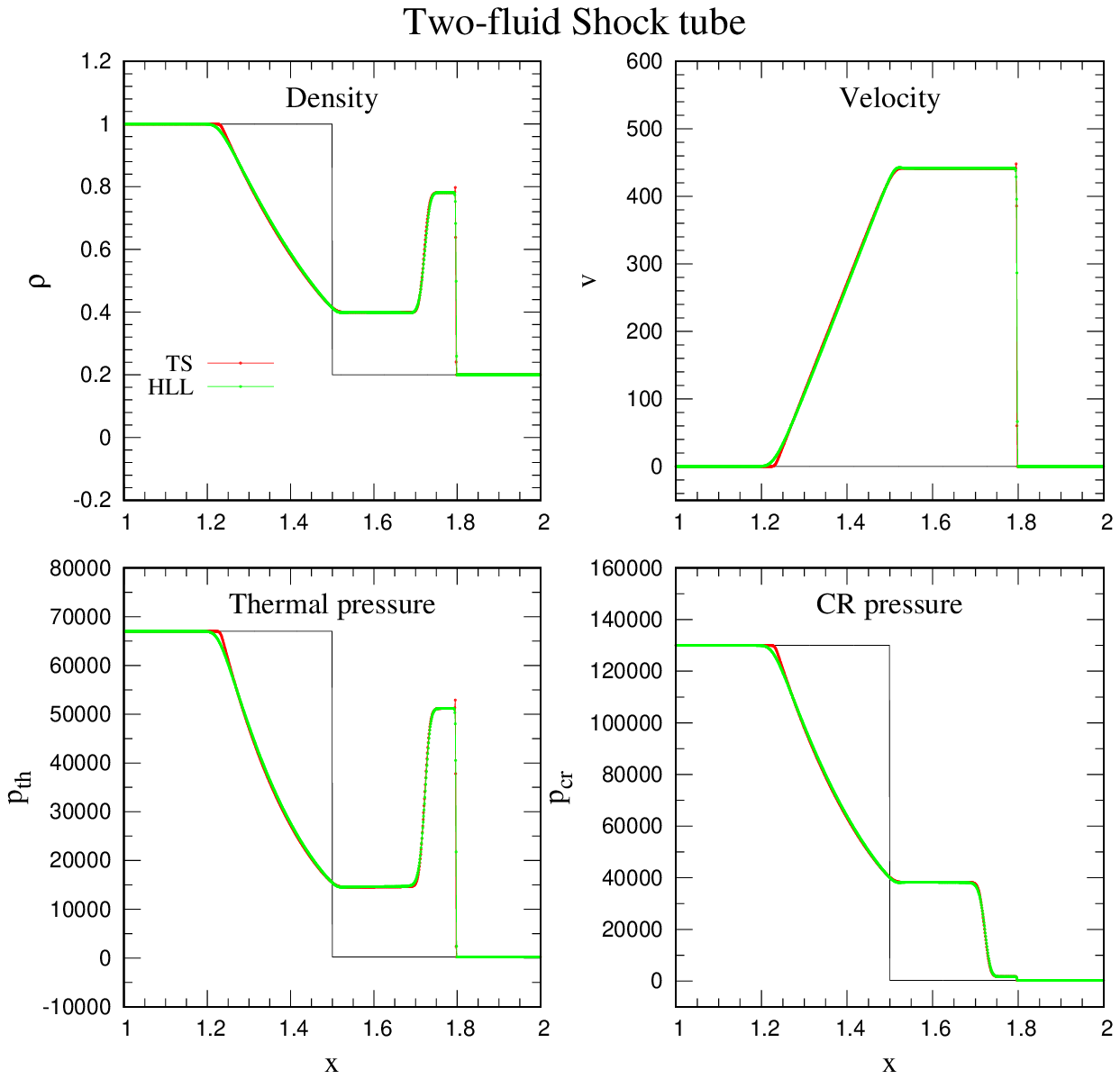}
\end{minipage}
\caption{Shock tube: a test problem in cartesian geometry. Left and right panels show the solutions of one-fluid and two-fluid shock-tube. Black lines represent initial profiles, blue colours stand for PLUTO, red and green colours represent the results of solvers `TS' and `HLL' solver of the new code (described in section \ref{sec:setup}). Figure shows that the results are quite same. For setup details, see section \ref{subapp:1dshk}.}
\label{fig:1dshk}
\end{figure*}
\begin{figure*}
\begin{minipage}[b]{0.45\linewidth}
\centering
\includegraphics[height=2.8in,width=2.9in]{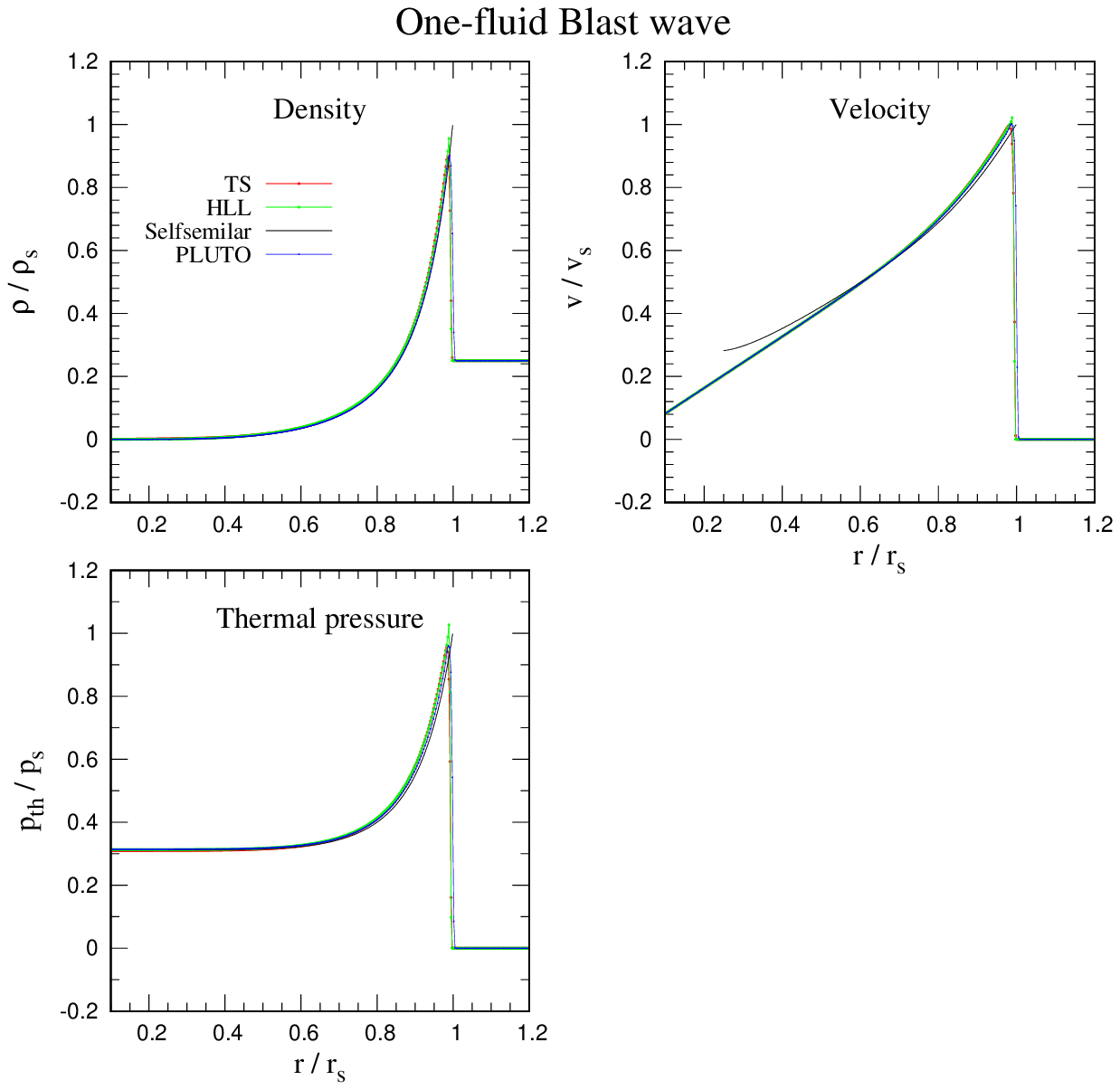}
\end{minipage}
\hspace{0.05cm}
\begin{minipage}[b]{0.45\linewidth}
\centering
\includegraphics[height=2.8in,width=2.9in]{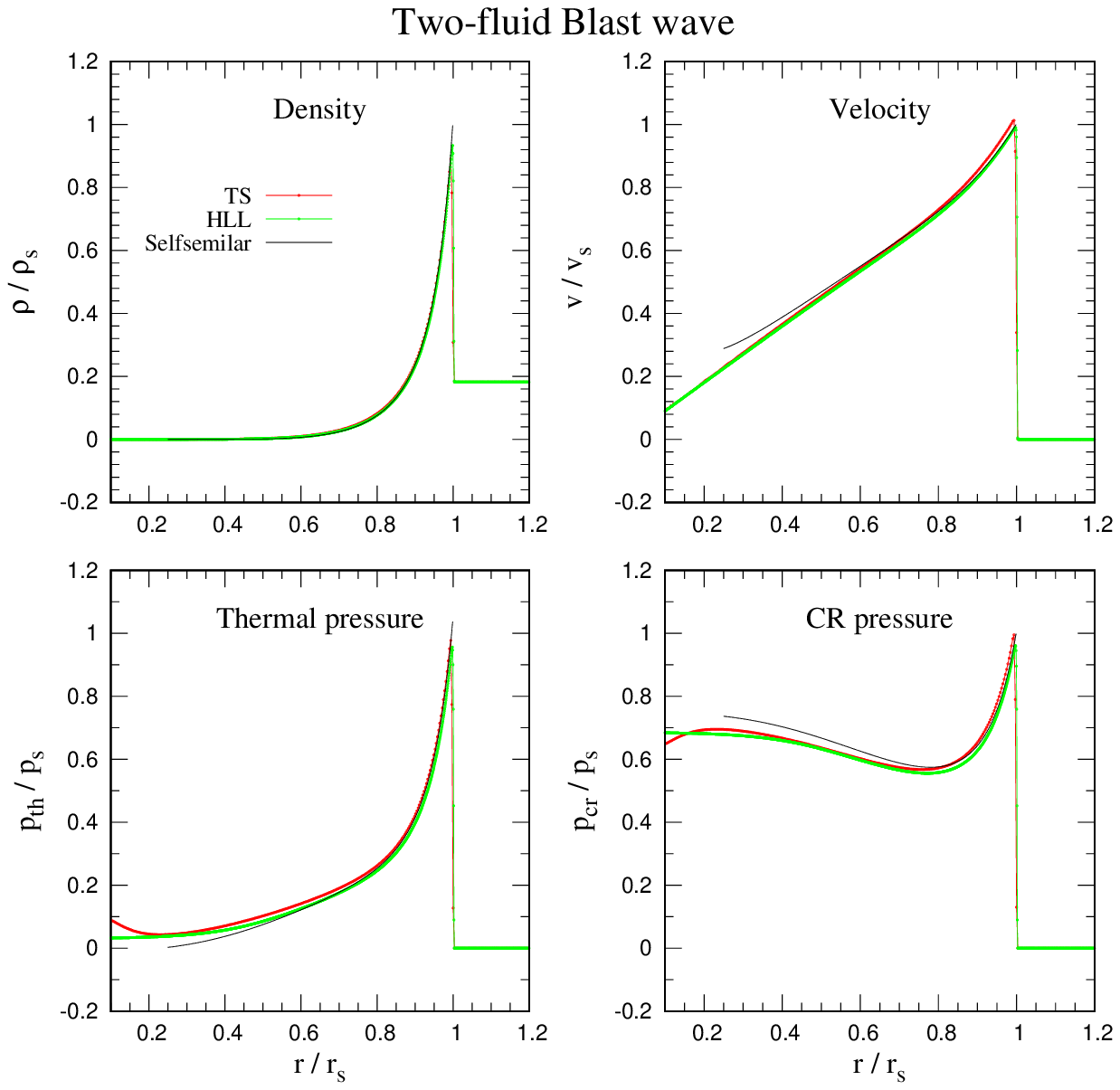}
\end{minipage}
\caption{Blast wave: a test problem in spherical geometry. All quantities are normalized w.r.t the self-similar variables. Left/right panels shows the one-fluid/two-fluid blast wave. Black colours display the solutions obtained from ODEs, blue colours stand for PLUTO, red and green colours stand for `TS' and `HLL' solvers of our code. The numerical solutions match quite well with the analytical results.}
\label{fig:blast}
\end{figure*}
\subsection{Shock-tube}
\label{subapp:1dshk}
This test problem is identical to the problem described in \citet{Sharma2013} (also see \citealt{Sod1978}; \citealt{Pfrommer2006}).
{\bf\noindent Problem set-up :} We set geometry to cartesian coordinate and choose total $2000$ grid points in a domain $[1,2]$. As the initial condition, the left state ($x<1.5$) is defined as $(\rho, v, p_{\rm th}, p_{\rm cr})_{\rm L} = (1, 0, 6.7\times10^4, 1.3\times10^5)$. The right state ($x\geq 1.5$) is defined as $(\rho, v, p_{\rm th}, p_{\rm cr})_{\rm R} = (0.2, 0, 2.4\times10^2, 2.4\times10^2)$. The CFL number is set to $0.4$. The snap shot of various profiles are shown in Figure \ref{fig:1dshk}. The left panels display the profiles for one fluid shock at $t=1\times 10^{-3}$ and the right panels display the profiles for two-fluid shock at $t=5\times 10^{-4}$.
\subsection{Blast wave}
\label{subappn:blast}
{\bf\noindent Problem set-up :} We set geometry to spherical coordinate and choose total $1000$ grid points in a domain $[0.1,4]$ pc. The initial profiles are: $\rho =1\,m{\rm _{H}\,cm^{-3}}$ (uniform), $v = 0.0$, $p_{\rm th} = (\rho/\mu m_{\rm H})k_{\rm B} T$ where $\mu=0.6$ and $T=10^4$ K. At $t=0$, at the first computational zone (say, volume $\delta V_{\rm src}$), we set energy density $10^{51}/\delta V_{\rm src}$. The CFL number is set to $0.2$. The snap shot of various profiles are shown in Figure \ref{fig:blast}. Left and right panels display the profiles of one fluid and two-fluid blast wave where all variables are scaled w.r.t to the self-similar variables (\citealt{Chevalier1983}). For the two-fluid run, first, we  identify the shock and set the CR pressure fraction to $w=1/2$ (see equation (\ref{eq:pcrin})). The profiles match quite well with the ODEs results (shown by black curves).
\subsection{Diffusion module}
\label{subappn:diffusion}
One useful test problem to check diffusion module was proposed by \citet{Reale1995}.\\
{\bf\noindent Problem set-up :} Recall their equations ($15$) and ($16$). We set $n=5/2$ , $a=4.412$ and $Q=1.2\times 10^{15}$ K cm. The length and temperature units are defined as $10^{8}$ cm and $\simeq 6.057\times10^{7}$ K. Total $400$ grid points are uniformly set in a domain [$0,5$]. The initial time is chosen as $t=0.1$ and the simulation is run upto $t=3.1$. For a detailed setup, see $\rm PLUTO/Testproblem/MHD/Thermal\_Conduction/TCfont$ (\citealt{Mignone2007}). We turn on only the diffusion module and our results are shown in Figure \ref{fig:testdiffusion}. For cartesian coordinate, we have compared the results with the analytical solution (black curves in left panel) given in equation ($16$) of \citet{Reale1995}. The comparison between PLUTO and \textsc{TFH} is also shown in Figure \ref{fig:testdiffusion}.
\begin{figure}
\centering
\includegraphics[height=1.75in,width=3.4in]{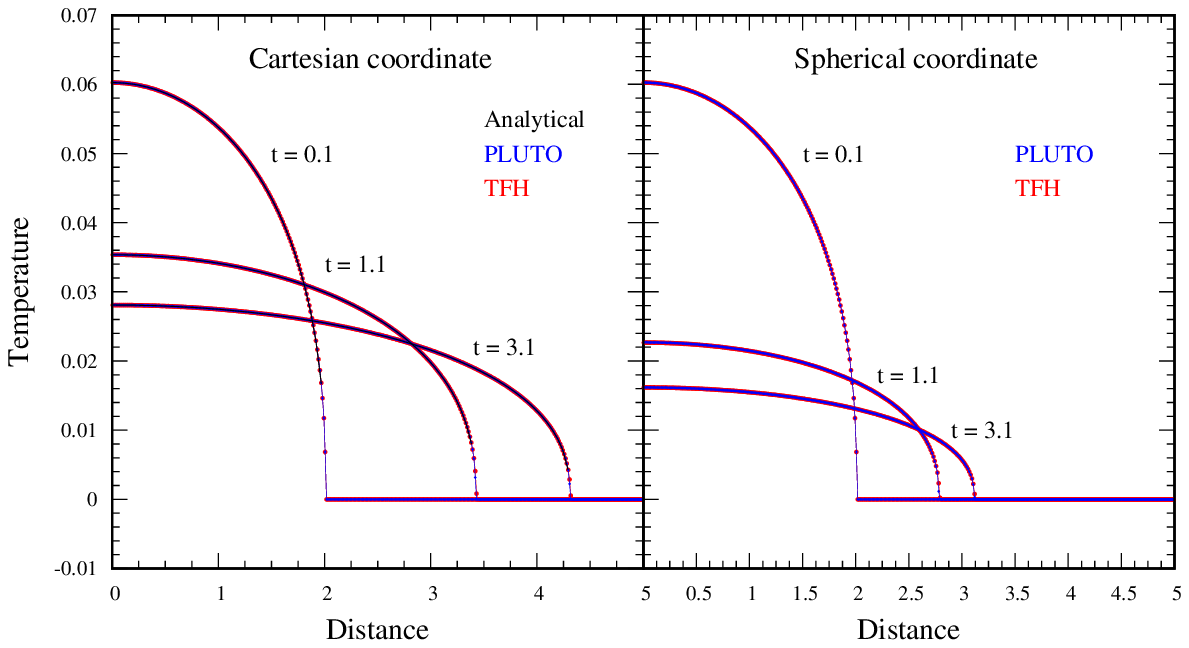} 
\caption{A test problem of diffusion module in cartesian (left panel) and spherical (right panel) geometry. The black curves in left panel denote the analytical results (see equation ($16$) in \citealt{Reale1995}). The blue and red colours represent \textsc{PLUTO} and \textsc{TFH} output respectively.}
\label{fig:testdiffusion}
\end{figure}
\section{Solver selection}
\label{app:solver}
TFH has two solvers (1) TS and (2) HLL (see section \ref{sec:setup}). Here we show a comparison of density profile of an adiabatic one-fluid ISB at $t=0.5$ Myr between HLL/TS solver of TFH and HLL solver of PLUTO in Figure \ref{fig:compISB}. The set-up is identical for all cases. This figure shows that 1st order HLL solver is more diffusive than 1st order TS solver explaining the reason for selecting `TS' solver.
\begin{figure}
\centering
\includegraphics[height=2.7in,width=2.7in]{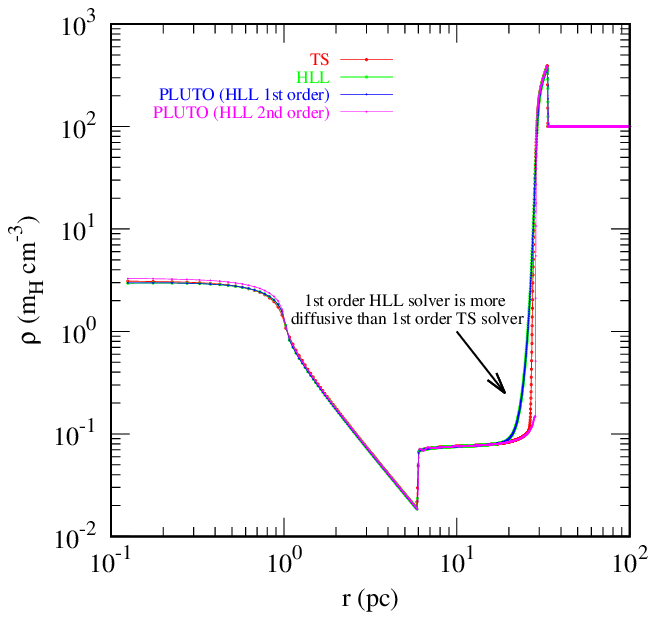} 
\caption{Comparison of ISB profiles at $0.5$ Myr between PLUTO and the new code. Blue and magenta colours stand for PLUTO, red and green colours stand for `TS' and `HLL' solvers of our new code respectively. Figure highlights the reason for choosing the `TS' solver.}
\label{fig:compISB}
\end{figure}
\section{Resolution test}
\label{app:resolution}
Here we present the convergent tests of a two-fluid ISB by considering four different spatial resolutions: $\Delta r = 0.1, 0.05, 0.025,0.0125$ pc. Figure \ref{fig:resol} displays the thermal and CR pressure profiles of an adiabatic two-fluid ISB (with CR diffusion) at $t_{\rm dyn}=0.9$ Myr and $t_{\rm dyn}=2.1$ Myr. The blue colours in both panels denote our fiducial resolution. This figure displays that the results are converged only for high resolutions ($\Delta r < 0.1$ pc). We have found that, the time when $P_{\rm cr} > P_{\rm th}$ increases with the decrease in spatial resolution (drastically when $\Delta r\gtrsim 0.2 $ pc). Resolution test for a more realistic bubble (i.e.; in addition to CR diffusion, cooling is on) is shown in Figure \ref{fig:resol2}. For our fiducial choice (resolution $\Delta r = 0.05$ pc), the results are well converged and the conclusions remain same.
\begin{figure}
\centering
\includegraphics[height=1.8in,width=3.4in]{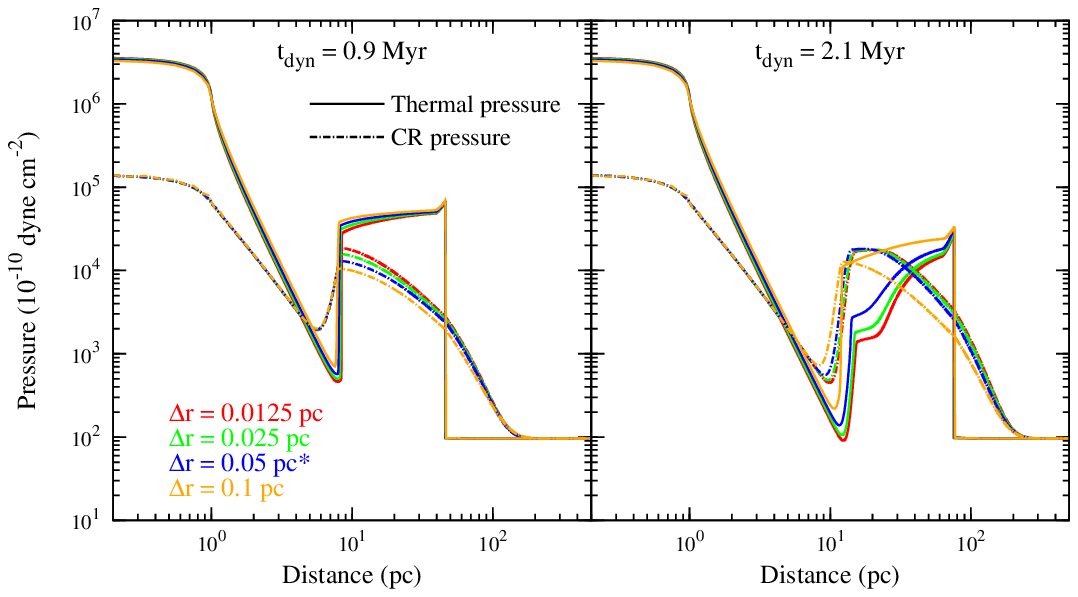} 
\caption{Resolution test. The solid lines represent thermal pressure and the dashed-dotted lines represent CR pressure profile. Left and right panels stands for two different dynamical time ($t_{\rm dyn}=0.9,2.1$ Myr). For both panels, the colour code is the same. The blue colour stands for our fiducial resolution. Figure highlights that the low resolution runs can take longer time to become CR dominated.}
\label{fig:resol}
\end{figure}

\begin{figure}
\centering
\includegraphics[height=1.8in,width=3.4in]{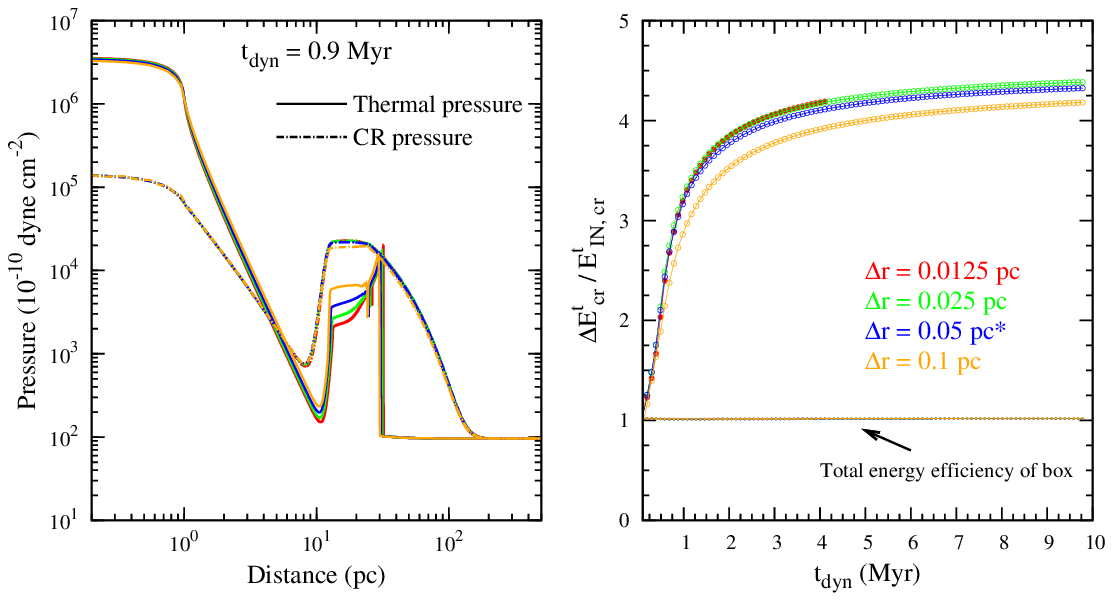} 
\caption{Resolution test for a two-fluid ISB with radiative cooling plus CR diffusion. Left panel shows CR pressure (dashed-dotted lines) and thermal pressure (solid lines) profiles at $t_{\rm dyn}=0.9$ Myr. Right panel shows the net change of CR energy as a function of $t_{\rm dyn}$ (also see Figure \ref{fig:fracgaincr}). The box size is taken as $r_{\rm max}=1000.1$ pc, except for the red curve where $r_{\rm max}=500.1$ pc and $t_{\rm dyn, end}=4.1$ Myr. The colour code is the same for both panels where blue colour represent our fiducial resolution.}
\label{fig:resol2}
\end{figure}
\end{document}